\documentclass[twocolumn]{aastex631}

\usepackage{comment}

\newcommand{\Msun}{{\rm M_{\odot}}}

\newcommand{\kpc}{\, {\rm kpc}}

\newcommand{\kmps}{\, {\rm km \, s^{-1}}}

\begin{document}

\title{Dynamically-Driven Evolution of Molecular Gas in the Barred Spiral Galaxy M83 \\
Traced by CO $J=$2-1/1-0 Line Ratio Variations}

\author[0000-0002-8762-7863]{Jin Koda}
\affiliation{Stony Brook University, Stony Brook, NY 11743-3800, USA}

\author[0000-0002-1639-1515]{Fumi Egusa}
\affiliation{Institute of Astronomy, Graduate School of Science, The University of Tokyo, 2-21-1 Osawa, Mitaka, Tokyo 181-0015, Japan}

\author[0000-0002-0465-5421]{Akihiko Hirota}
\affiliation{Joint ALMA Observatory, Alonso de C\'ordova 3107, Vitacura, Santiago 763-0355, Chile}
\affiliation{National Astronomical Observatory of Japan, 2-21-1 Osawa, Mitaka, Tokyo 181-8588, Japan}

\author[0000-0001-8254-6768]{Amanda M Lee}
\affiliation{Stony Brook University, Stony Brook, NY 11743-3800, USA}

\author[0000-0002-0588-5595]{Tsuyoshi Sawada}
\affiliation{Joint ALMA Observatory, Alonso de C\'ordova 3107, Vitacura, Santiago 763-0355, Chile}
\affiliation{National Astronomical Observatory of Japan, 2-21-1 Osawa, Mitaka, Tokyo 181-8588, Japan}

\author[0000-0002-8868-1255]{Fumiya Maeda}
\affiliation{Research Center for Physics and Mathematics, Osaka Electro-Communication University, 18-8 Hatsucho, Neyagawa, Osaka, 572-8530,  Japan}

\shorttitle{CO 2-1/1-0 across M83}
\shortauthors{Koda et al.}

\begin{abstract}
We show the variations of the CO $J=$2-1/1-0 line ratio ($R_{21/10}$) across the barred spiral galaxy M83,
using the 46~pc resolution data from the Atacama Large Millimeter/submillimeter Array (ALMA).
The $R_{21/10}$ map clearly evidences the systematic large-scale variations as a function of galactic structures.
Azimuthally, it starts from low $R_{21/10}\lesssim 0.7$ in the interarm regions and becomes high $\gtrsim 0.7$ in the bar and spiral arms, suggesting that the density and/or kinetic temperature of molecular gas increase by about a factor of 2-3.
This evolution is seen even in the parts of spiral arms without star formation,
and $R_{21/10}$ is often elevated even higher to $\sim$0.8-1.0 when HII regions exist in the vicinity.
Radially, $R_{21/10}$ starts very high $\gtrsim$1.0 at the galactic center, remains low $\lesssim$0.7 in the bar region, increases to $\gtrsim$0.7 around the bar end, and again decreases to $\lesssim$0.7 in the rest of disk where the spiral arms dominate.
The evolutionary sequence is synchronized with galactic rotation, and therefore, it is determined largely by the galactic structures and dynamics and is governed by the galactic rotation timescales.
The $R_{21/10}$ map also shows that the influence of stellar feedback is localized and limited.
Massive, large, and non-star forming molecular structures have low $R_{21/10}$,
which also suggests that the bulk molecular gas in the disk is not regulated by stellar feedback, but more likely by galactic structures and dynamics.
These results are consistent with suggestions by the earlier studies of the Milky Way and other barred spiral galaxies, and thus, are likely general among barred spiral galaxies in the local Universe.
\end{abstract}

\section{Introduction} \label{sec:intro}

The evolution of molecular gas and clouds is a key step to understand star formation (SF) and galaxy evolution.
Virtually all SF occurs in the molecular gas and clouds.
Galactic structures and dynamics can perturb the gas clouds and stimulate their evolution \citep[e.g, ][]{Scoville:1979lg, Vogel:1984aa, Aalto:1995lr, Koda:2009wd}.
The subsequent SF and feedback are important in galaxy growth or quenching.
These physical processes should leave imprints on the distribution of molecular gas and clouds and their physical conditions.
In this paper, we show that the evolution of molecular gas and clouds is synchronized with large-scale galactic structures, using the CO $J=$2-1/1-0 line ratio ($R_{21/10}$).

$R_{21/10}$ is an important diagnostic tracer of the physical conditions of molecular gas \citep{Goldreich:1974ab, Scoville:1974yu}.
It is sensitive to even mild changes in the kinetic temperature $T_{\rm k}$ and/or H$_2$ volume density $n_{\rm H_2}$, as well as the CO column density $N_{\rm CO}$.
For example, $R_{21/10}$ could increase with the gas compression toward spiral arms (increase in $n_{\rm H_2}$ before SF) or heating from stellar feedback (increase in $T_{\rm k}$ after SF)  \citep{Koda:2012lr}.
It could decrease as the gas passes the spiral arms and expands into the interarm regions.
The CO 1-0 and 2-1 emissions are the only two molecular lines significantly detectable over the whole disks of nearby galaxies with current facilities.
Therefore, $R_{21/10}$ is the most viable tool for mapping the physical state of the bulk molecular gas across galactic disks.

A series of seminal work by the 60~cm telescopes at the Nobeyama Radio Observatory and La Silla Observatory showed the systematic variations of $R_{21/10}$ within/between local molecular clouds and across the Milky Way (MW) disk \citep[][for review]{Hasegawa:1997lr}.
For the line ratio analyses, they built 60~cm-diameter telescopes dedicated for CO(2-1), whose beam sizes match with those of the 1.2~m telescopes for the Galactic plane CO(1-0) survey \citep{Dame:1987aa, Dame:2001gs}.

They found that $R_{21/10}$ changes systematically from $>$0.7 (\textit{High Ratio Gas} - HRG) to $<$0.7 (\textit{Low Ratio Gas} - LRG) between star-forming and dormant molecular clouds \citep{Sakamoto:1994aa, Sakamoto:1997fk}, between spiral arms and interarm regions \citep{Sakamoto:1997fk, Yoda:2010rf}, and from the galaxy center to the outskirts \citep{Oka:1996lr, Oka:1998fk, Sawada:2001lr}.
$R_{21/10}>1$ (\textit{Very High Ratio Gas} - VHRG) rarely occurs except in the immediate vicinity of on-going SF \citep{Sakamoto:1994aa}, in the small shock surfaces of supernova explosions \citep{Seta:1998aa}, or in the Galactic center \citep{Oka:1996lr}.
The average $R_{21/10}$ is as high as 0.96 in the central 900~pc diameter region of the Galactic center and is 0.6-0.7 in the Galactic disk \citep{Sawada:2001lr}.
From the distribution and size of HRG and VHRG as well as the deficiency of atomic gas in the inner Galaxy \citep[e.g., ][]{Nakanishi:2016aa, Koda:2016aa},
\citet{Sakamoto:1997fk} suggested that the increase of $R_{21/10}$ from LRG to HRG occurs due predominantly to the effects of galactic structures and dynamics, not much to stellar feedback.

Many efforts have been made to characterize spatial variations of $R_{21/10}$ within external galaxies,
but encountered difficulties in data calibration \citep[e.g., ][]{Crosthwaite:2002yu, Lundgren:2004aa, Crosthwaite:2007uq, Yajima:2021aa, den-Brok:2021aa, Leroy:2022ac}.
They often found $R_{21/10}>1$ over very large areas in galaxies, which is implausible given that the required physical conditions are difficult to achieve over the large areas (i.e., optically-thin, but high $n_{\rm H_2}$ and $T_{\rm k}$).
Only more dedicated studies with emphasis on calibrations were able to show the large-scale $R_{21/10}$ variations before the ALMA era \citep{Koda:2012lr, Vlahakis:2013aa}.
Such studies are now becoming possible with new facilities, especially ALMA \citep{Koda:2020aa, Maeda:2022aa, Egusa:2022aa, den-Brok:2023ab, den-Brok:2024aa}.
However, the previous ALMA studies did not explore the full ALMA capability and are limited by sensitivity and resolution due to relatively short observations.
We also note that the spatially-unresolved $R_{21/10}$ surveys of galaxies, often with one pointing with a single dish telescope, were also performed \citep[e.g., ][]{Braine:1992lr, Saintonge:2017aa, Keenan:2024aa}.

In this paper, we show the $R_{21/10}$ variations across the barred spiral galaxy M83.
This galaxy shows a close morphological resemblance to the MW and is an ideal target for comparisons with the previous MW results \citep[see ][for their parameters]{Koda:2023aa}.
Its proximity ($d=$4.5~Mpc) permits us to achieve a high mass sensitivity and linear resolution, which is demonstrated by the recent high-fidelity CO(1-0) imaging with ALMA \citep{Koda:2023aa}.
This paper discusses the large-scale distribution and then the dormant, non-star forming molecular gas across the galaxy.
Impacts of SF will be discussed in depth in a separate paper.

\section{Data} \label{sec:reduction}

\begin{figure*}[th]
\centering
\includegraphics[width=1.0\textwidth]{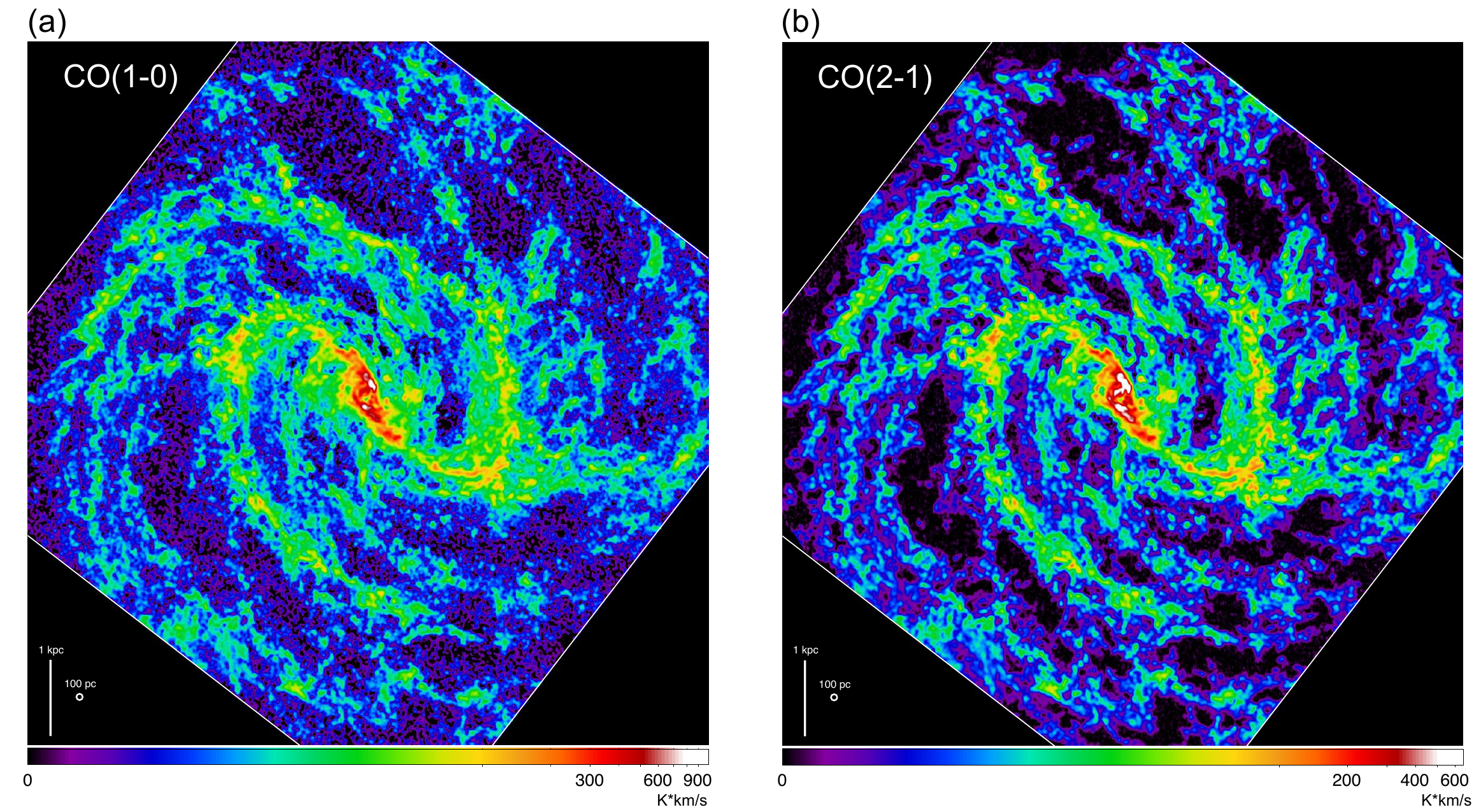}
\caption{Integrated intensity maps in CO(1-0) and CO(2-1), $I_{10}$ and $I_{21}$, respectively.
For a fair comparison of the two transitions, the plot ranges are adjusted to show prominent structures roughly with the same colors, by assuming a typical $R_{21/10}(I)$ of $\sim 0.65$.
The CO(1-0) map shows more extended emission than the CO(2-1) map.
}
\label{fig:co10co21moms}
\end{figure*}

We use the CO(1-0) and CO(2-1) data obtained by the ALMA's 12m, 7m, and Total Power (TP) arrays:
\#2017.1.00079.S for CO(1-0), and \#2013.1.01161.S, 2015.1.00121.S \& 2016.1.00386.S for CO (2-1).
The methods of data reduction and imaging were discussed in depth by \citet{Koda:2020aa, Koda:2023aa} for the TP and interferometer data, respectively.
The data are calibrated with the Common Astronomy Software Application 
\citep[CASA; ][]{CASA-Team:2022aa}.
The TP cube is converted to visibilities using the Total Power to Visibilities (TP2VIS) package \citep{Koda:2019aa}.
The 12m, 7m, and TP data are jointly imaged with the Multichannel Image Reconstruction, Image Analysis, and Display (MIRIAD) package \citep{Sault:1995kl, Sault:1996uq}.
The CLEAN algorithm in MIRIAD takes into account the spatially-variable point spread function, which is important for the large-scale mosaic imaging \citep[see ][ for more discussion]{Koda:2023aa}.

We have improved the imaging process since \citet{Koda:2023aa}.
The details of the data reduction and imaging are discussed in the papers mentioned above with some additional information in Appendix \ref{sec:reductiondetails}.
In short, we CLEANed the data much deeper with a mask that encloses the emission to a very faint level.
We repeated the CLEAN imaging three times and improved the mask each time.
This iterative procedure and very deep CLEANing removed low-level negative sidelobes.

For both CO(1-0) and CO(2-1) data, we adopted a round restoring beam of 2.1" (46~pc) in the imaging.
The data cubes have the pixel size of 0.25" in RA and DEC and $5\kmps$ in velocity.
The root-mean-square (RMS) noises are 65 and 25~mK in a $5\kmps$ channel in CO(1-0) and CO(2-1), respectively.
Therefore, the molecular gas with $R_{21/10}\sim$0.4 is detected at around the same significance in CO(1-0) and CO(2-1).

\subsection{Maps} \label{sec:moments}

Figure \ref{fig:co10co21moms} shows the CO(1-0) and CO(2-1) integrated intensity maps, $I_{\rm 10}$ and $I_{\rm 21}$, respectively.
To generate these maps, we applied a mask to the data cubes.
We used a signal-to-noise (S/N) ratio cube of CO(1-0) and made the mask by selecting the volumes (pixels) with at least one pixel with $>3.5\sigma$ and more than 32 pixels with $>2.5\sigma$, and extended their volumes to the pixels with $>1.5\sigma$.
The choice of these thresholds is arbitrary and is set by trial and error.
This mask was expanded by 20" (80 pixels) in all spatial directions to capture surrounding diffuse emission below the cutoff of $1.5\sigma$ in case there is any.
The emission was integrated along velocity within the mask.
The corresponding error maps, $\Delta I_{10}$ and $\Delta I_{21}$, were also generated.
$I_{21}$ is typically about 65\% of $I_{10}$, and thus, for a fair comparison,
they are plotted in a range of $I_{10}\in [0, 1000]$ and $I_{21}\in [0, 650]$ in units of ${\rm K\cdot km/s}$.

\begin{figure*}[t!]
\centering
\includegraphics[width=1.0\textwidth]{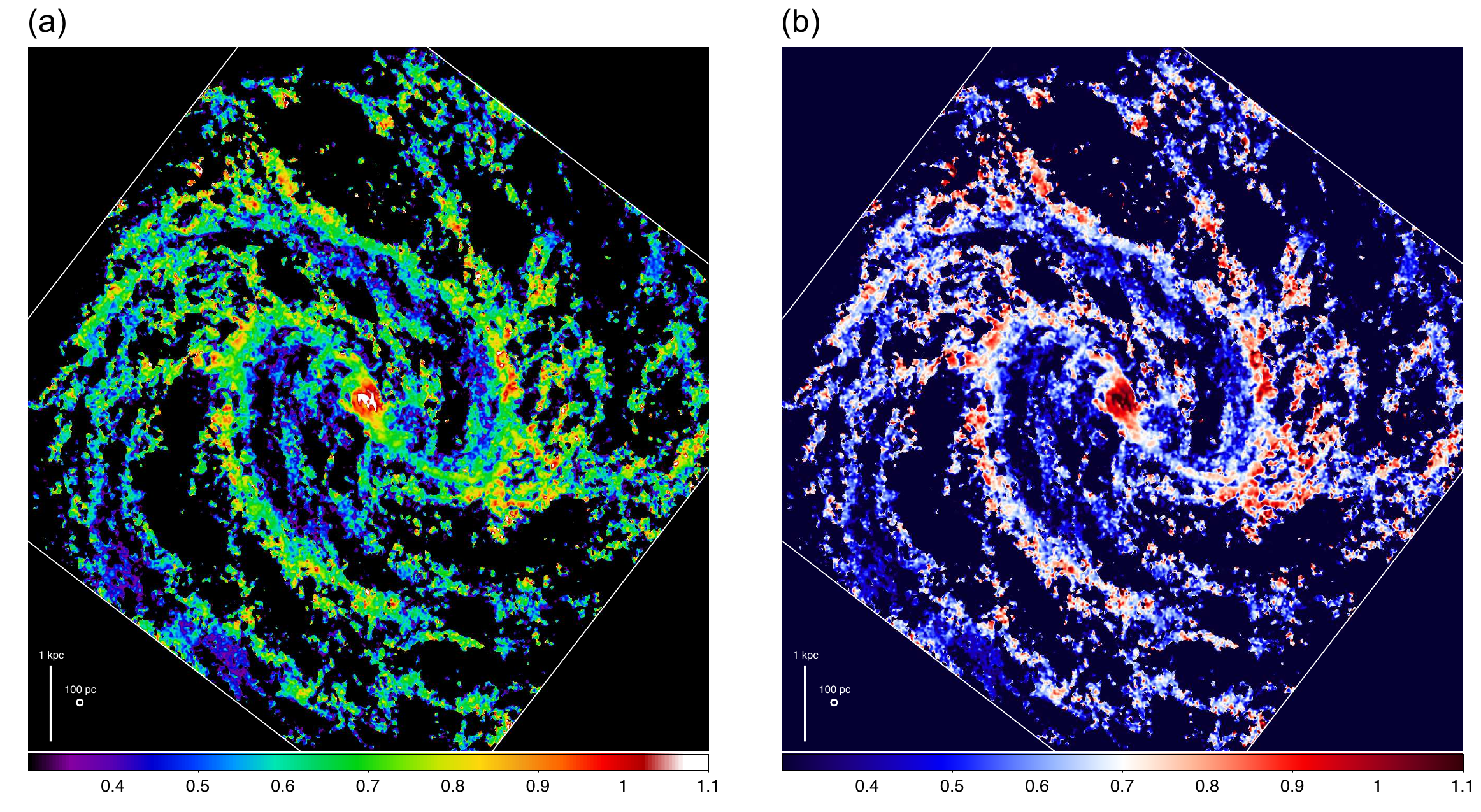}
\caption{The $R_{21/10}(I)$ map with two color palettes.
The right panel has the dividing line of $R_{21/10}\sim 0.7$ in white.
The high ratio gas ($>0.7$) is red, and the low ratio gas ($<0.7$) is blue.
The left panel uses a wider range of color to present $R_{21/10}$ at a finer scale.
}
\label{fig:R21map}
\end{figure*}

Figure \ref{fig:R21map} shows the CO(2-1)-to-CO(1-0) line ratio map $R_{21/10}(I) \equiv I_{21}/I_{10}$.
Only the pixels with $I_{10}/\Delta I_{10}>3$ and $I_{21}/\Delta I_{21}>3$ are used for this map.
These sensitivity-based criteria are relatively lenient, but are sufficient for the discussions of this paper.
The effects of this sensitivity cut are discussed further in Appendix \ref{sec:senscut}.

We also generated the CO(1-0) and CO(2-1) peak brightness temperature maps, $T_{10}$ and $T_{21}$, respectively.
Along the spectrum in each spatial pixel, the peak brightness in CO(1-0) was taken for these maps.
The CO(2-1) brightness at the same velocity was used for CO(2-1).
The $R_{21/10}$ defined with the peak brightness, $R_{21/10}(T) \equiv T_{21}/T_{10}$, is used only occasionally in this paper (Section \ref{sec:R21overall}).

We denote $R_{21/10}(I)$ as $R_{21/10}$ for simplicity.
$R_{21/10}(T)$ is always expressed as $R_{21/10}(T)$.

\section{Classifications on $R_{21/10}$} \label{sec:classification}

The U. Tokyo group classified the molecular gas across the Galactic disk according to $R_{21/10}$ as: \textit{Low Ratio Gas} (LRG: $<0.7$),
\textit{High Ratio Gas} (HRG: =0.7-1.0), and
\textit{Very High Ratio Gas} (VHRG: $>1$) \citep{Hasegawa:1997lr, Sakamoto:1997fk, Oka:1998fk}.
Their classification scheme is motivated by observational and theoretical considerations.
Their Galactic studies clearly laid out the scopes and limitations of $R_{21/10}$ analyses.
Therefore, we adopt their classifications, in particular, their dividing lines of $R_{21/10}=$0.7 and 1.0.
In addition, we occasionally use a subclass in HRG, $R_{21/10}=$0.8-1.0, as $R_{21/10}$ around HII regions is often elevated not only to $>0.7$ but even higher to $>0.8$ (Sections \ref{sec:relation2SF} and \ref{sec:noSFgas}).

This section discusses these dividing lines.

\subsection{$R_{21/10}=$0.7 Between HRG and LRG} \label{sec:R21_0.7}

The first dividing line of $R_{21/10}\sim$0.7 is partly from the theoretical calculations with the Large Velocity Gradient (LVG) model \citep{Goldreich:1974ab, Scoville:1974yu, van-der-Tak:2007qy}.
It predicts $R_{21/10}\sim$0.6-0.7
for a typical Galactic molecular cloud of ($n_{\rm H_2}$, $T_{\rm k}$)$\sim$($300{\,\rm cm^{-3}}$, $10{\,\rm K}$) \citep{Scoville:1987vo, Koda:2023aa}.
This is also around the observed average in the Galactic disk \citep[$\sim$0.6-0.7; ][]{Sakamoto:1997fk, Sawada:2001lr}.
Simply put, HRG is denser and warmer than the average molecular gas in the Galactic disk.
LRG is less dense and cooler than the average. 
The transition from LRG to HRG (or the reverse) is continuous in the values of $n_{\rm H_2}$ and $T_{\rm k}$ (and $N_{\rm CO}$).
Nevertheless, as a crude guideline, HRG has about 2-3 times higher $n_{\rm H_2}$ and $T_{\rm k}$ than LRG according to the LVG calculation by \citet[][see their Figure 7 for the ranges and limitations, as well as numerous LVG studies in the literature, e.g., \citealt{Scoville:1974yu, Goldsmith:1983aa, Sakamoto:1994aa, Oka:1998fk}]{Koda:2012lr}.

$R_{21/10}$ is not uniform within individual molecular clouds.
An average $R_{21/10}$ of each cloud is determined by the mixing fraction of HRG and LRG \citep{Sakamoto:1997fk}.
Indeed, local clouds with massive SF (e.g., Orion) have a larger fraction of HRG, and thus, have a larger average $R_{21/10}$ \citep{Sakamoto:1994aa}.
Dormant non-star forming clouds have a larger fraction of LRG and show a lower $R_{21/10}$ \citep{Sakamoto:1997fk}.
The variations in the mixing fraction cause systematic variations in $R_{21/10}$: from $>$0.7 to $<$0.7 between star-forming and dormant molecular clouds \citep{Sakamoto:1994aa, Sakamoto:1997fk}, between spiral arms and interarm regions \citep{Sakamoto:1997fk, Yoda:2010rf}, and from the galaxy center to the outskirts \citep{Oka:1996lr, Oka:1998fk, Sawada:2001lr}.
The average becomes as high as $R_{21/10}=$0.96 in the central 900~pc region of the MW \citep{Sawada:2001lr}.

We occasionally mention a sub-range of HRG, $R_{21/10}=$0.8-1.0.
This is empirically motivated as $R_{21/10}$ is often elevated not only to $>0.7$ but even higher to $>0.8$ around HII regions (Sections \ref{sec:relation2SF} and \ref{sec:noSFgas}).

\subsection{$R_{21/10}>$1.0 for VHRG} \label{sec:R21_1.0}

VHRG ($R_{21/10}>1$) does not occur easily because it has to satisfy seemingly contradicting conditions.
The gas has to remain optical-thin ($\approx$ low $N_{\rm CO}$) when it needs higher $n_{\rm H_2}$ and $T_{\rm k}$ for the very high excitation.
Obviously, when they are optically-thick, both the CO(1-0) and CO(2-1) brightness temperatures approach $T_{\rm k}$ (minus the Cosmic Microwave Background temperature) based on the basic radiative transfer equation. 
Hence, $R_{21/10}$ converges to unity.
VHRG is observed in only a part of molecular clouds under a direct influence of on-going SF \citep{Sakamoto:1994aa}, in small shocked regions (small pockets) at the edges of supernova remnants \citep{Seta:1998aa}, and in the Galactic center \citep{Oka:1996lr}.

\section{Large-Scale Variations of $R_{21/10}$} \label{sec:R21}

We discuss $R_{21/10}$ and its large-scale variations using the classifications with the main dividing lines of $R_{21/10}=$0.7 and 1.0 (see Section \ref{sec:classification}).
We will occasionally use $R_{21/10}=$0.8-1.0 as an additional subclass, as it often occurs when the gas is interacting with star-forming regions.

The $R_{21/10}$ values reported here represent the average ratios within individual molecular clouds.
The spatial resolution of 46~pc is close to the typical cloud diameter of 40~pc \citep{Scoville:1987vo}.
In fact, molecular clouds are identified in the CO(1-0) data \citep{Hirota:2024aa}.
When the clouds are isolated in each beam, the line ratio does not suffer from the beam filling factor.

\subsection{Distributions of CO(1-0) and CO(2-1) Emission} \label{sec:I10I21distribution}

In Figure \ref{fig:co10co21moms}, both CO(1-0) and CO(2-1) emissions are concentrated around the prominent galactic structures, such as the center, bar, and spiral arms in a similar manner.
However, CO(1-0) is more spatially extended, especially toward their surrounding regions (e.g., interarm regions).
This is expected as CO(2-1) requires higher excitation, and the gas outside the prominent structures are naturally less excited \citep[][]{Sakamoto:1997aa, Koda:2012lr, Koda:2020aa}.
There has been a notion that CO(1-0) and CO(2-1) are equivalent in tracing the amount of molecular gas,
and hence, that CO(2-1) can substitute CO(1-0) as a mass tracer \citep[e.g., ][]{Leroy:2008fj, Bigiel:2008aa, Leroy:2013aa, Sun:2018aa, Sun:2020tt, Leroy:2021ab}.
Figure \ref{fig:co10co21moms} shows that this notion has a limitation when galactic structures are resolved.

\subsection{Overall Variations of $R_{21/10}$} \label{sec:R21overall}

Figure \ref{fig:R21map} clearly shows that $R_{21/10}$ varies systematically with galactic structures in the radial and azimuthal directions.
The most evident is the azimuthal variations.
$R_{21/10}$ is low in the interarm regions ($\lesssim 0.7$) and is elevated in the bar ($\sim 0.7$) and spiral arms ($\gtrsim 0.7$).
Therefore, the $R_{21/10}$ map alone firmly evidences that the gas physical conditions are evolving in sync with large-scale galactic rotation, structures, and dynamics.
There is also the gas excited further to $R_{21/10}\sim$0.8-1.0, which is localized along the spiral arms and in the galactic center (see Sections \ref{sec:relation2SF} and \ref{sec:hst} for discussions on this component).
In the radial direction,
$R_{21/10}$ is high ($\sim$1.0) in the central region.
It goes down to $\lesssim 0.7$ in the bar region, becomes high $\gtrsim 0.7$ at the bar ends, and gradually decreases to $\lesssim 0.7$ in the disk where the spiral arms dominate (see also Section \ref{sec:radaz}).

\begin{figure}[b!]
\centering
\includegraphics[width=0.45\textwidth]{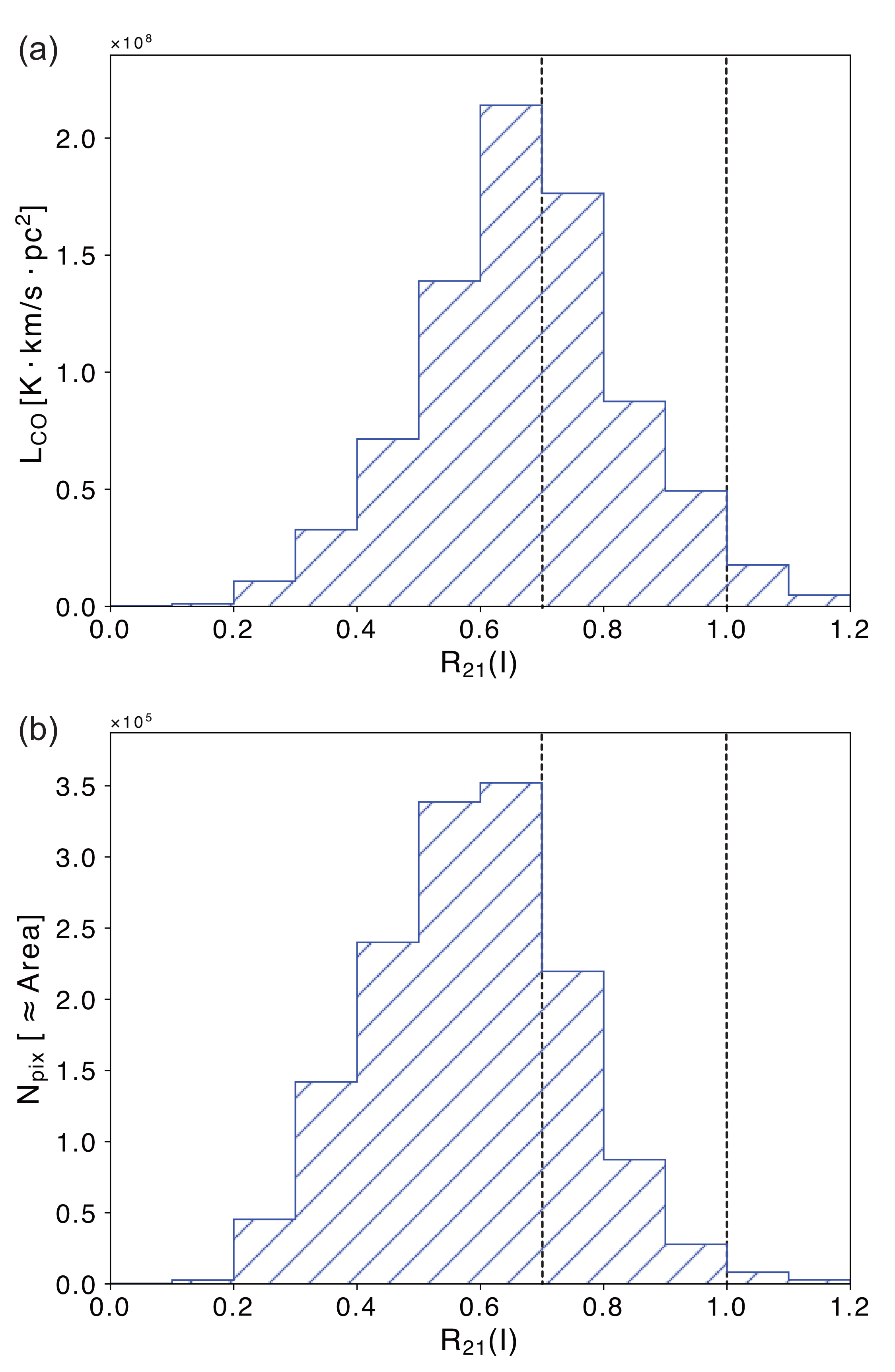}
\caption{Histograms of $R_{21/10}(I)$ over the whole disk,
(a) in CO(1-0) luminosity $L_{\rm CO}$, and
(b) in pixel counts $N_{\rm pix}$, which is equivalent to the area they occupy.
The vertical dotted lines show $R_{21/10}$=0.7 and 1.0 for reference.
}
\label{fig:R21hist_disk}
\end{figure}

Figure \ref{fig:R21hist_disk} shows the histograms of $R_{21/10}$ over the entire disk, (a) in flux and (b) in area (pixel counts).
In terms of area, the low ratio gas ($R_{21/10}<0.7$) spreads over large areas (76\% of the whole disk) compared to the high ratio gas ($>0.7$) that occupies 24\%.
The high ratio gas is typically brighter; and in flux, the low and high ratio gas encompass 58\% and 42\% of the total flux of the entire disk, respectively.
That is, the low and high ratio gas occupy a similar amount of flux, and thus, a similar amount of molecular gas mass if the CO-to-H$_2$ conversion factor $\alpha_{\rm CO}$ is constant between them.
[Note that the total H$_2$ mass involved in this analysis (in our mask) is $M_{\rm H_2}=2.6\times 10^9\Msun$ ($M_{\rm gas}=3.5\times 10^9\Msun$ including Helium and other heavier elements) for the MW conversion factor of $\alpha_{\rm CO}=4.35$, which is consistent with the total H$_2$ mass reported by \citet{Koda:2023aa}.
It becomes $M_{\rm H_2}=1.9\times 10^9\Msun$ for the recently-derived $\alpha_{\rm CO}=3.14$ in M83 by \citet{Lee:2024aa}.
We also note that $\alpha_{\rm CO}$ could change radially \citep{Lee:2024aa}, though it is beyond the scope of this paper.]

\begin{figure}[t!]
\centering
\includegraphics[width=0.48\textwidth]{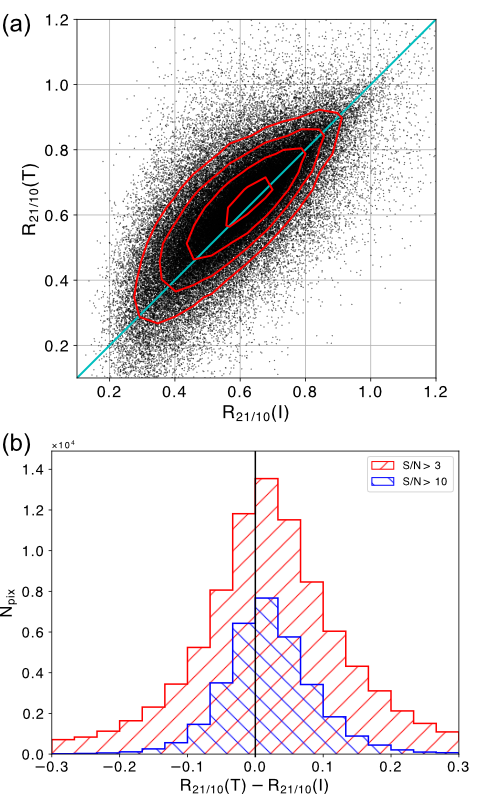}
\caption{Comparison of the CO(2-1)/CO(1-0) line ratio using the integrated intensity, $R_{21/10}(I) \equiv I_{21}/I_{10}$ and peak brightness temperature, $R_{21/10}(T) \equiv T_{21}/T_{10}$.
(a) $R_{21/10}(T)$-$R_{21/10}(I)$ plot.
Only every 4th spatial pixel in the RA and DEC directions is used.
The red contours are for the density of data points at 10, 20, 40, and 80\% of the peak density.
(b) Histogram of $R_{21/10}(T)-R_{21/10}(I)$ with the data of panel (a) (i.e., S/N$>$3), and with a more stringent sensitivity cut of $I_{10}/\Delta I_{10}>10$ and $I_{21}/\Delta I_{21}>10$ (i.e., S/N$>$10).
The distributions are skewed rightward.
}
\label{fig:R21mom0mom8}
\end{figure}

Figure \ref{fig:R21mom0mom8} compares the line ratios calculated in two ways, $R_{21/10}(I)$ (denoted as $R_{21/10}$ in this paper) and $R_{21/10}(T)$.
To reduce the data density in the plot, only every 4th spatial pixel (i.e., every 1") in the RA and DEC directions are plotted.
This roughly corresponds to the Nyquist sampling of the 2.1" beam size.
The correlation between $R_{21/10}(I)$ and $R_{21/10}(T)$ suggests that when the line peak location within a cloud is excited, the whole cloud tends to be excited.
At the same time, $R_{21/10}(T)$ is typically higher than $R_{21/10}(I)$, indicating that the high ratio gas is concentrated more around the velocity of the spectral line peak within individual clouds.
The line wings of the clouds have lower ratios, and thus, the average ratio over the clouds, $R_{21/10}(I)$, shows lower ratios.
This suggests that the $R_{21/10}$ changes within individual molecular clouds.

The observed range of $R_{21/10}$ is largely contained in $R_{21/10} \leq 1$ (Figure \ref{fig:R21hist_disk}), which gives us confidence in the data calibration and imaging.
A higher ratio of $R_{21/10}>1$ rarely occurs except in some localized regions, such as the galactic center (Figure \ref{fig:R21map}) and vicinities of star-forming regions (discussed later).
This makes sense from the astrophysical consideration (Section \ref{sec:R21_1.0}).
To have $R_{21/10}>1$, the gas has to be optically-thin (with high $n_{\rm H_2}$ and $T_{\rm k}$),
but the two transitions of CO are normally very optically-thick \citep[$\tau\sim$10-100; ][]{Goldreich:1974ab}.
In the Milky Way, the optically-thin condition is seen only in the regions of high energy (e.g., galactic center, supernova remnants, and vicinity of HII regions), which increases the gas velocity width (dispersion) and reduces the effective opacity per velocity.

\subsection{Radial and Azimuthal Variations} \label{sec:radaz}

\begin{figure}[h!]
\centering
\includegraphics[width=0.45\textwidth]{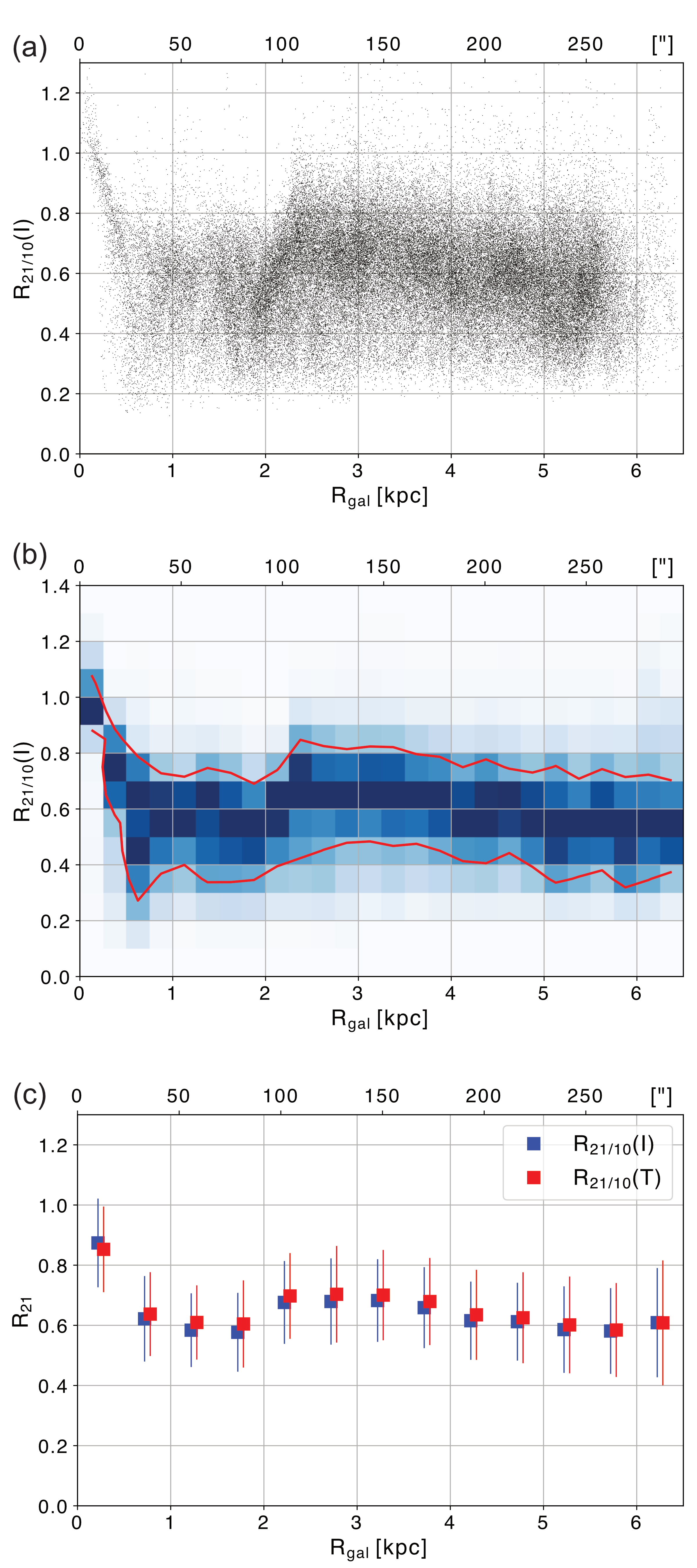}
\caption{
Radial distribution of $R_{21/10}$ with different presentations, (a) data points, (b) data density, and (c) CO(1-0)-intensity weighted average and standard deviations.
The two sets are shifted slightly in the horizontal direction for clarity of the presentation.
In panel (a), only every 4th spatial pixel in the RA and DEC directions are plotted to reduce the data density.
}
\label{fig:R21rad}
\end{figure}

Figure \ref{fig:R21rad} shows the $R_{21/10}$ variations along the galactic radius $R_{\rm gal}$ with three presentations: with
(a) data points,
(b) data density in shade and half-peak contours, and
(c) CO(1-0)-intensity weighted averages and standard deviations in radial bins.
This figure shows both radial and azimuthal variations (scatters) and also demonstrates that the azimuthal variations are larger.

In panel (a), every 4th pixel in the RA and DEC directions (i.e., 1/16th of the data in Figure \ref{fig:R21map}) are plotted.
In panel (b), the shade is normalized so that the peak density is 1 in each radial bin with a 0.25~kpc width.
The contours are drawn at 0.5 after the normalization.
In panel (c), the statistical values are calculated in each radial bin of a 0.5~kpc width (see Table \ref{tab:R21stat}).
The intensity-weighted average $R_{21/10}$ from $R_{21/10}(I)$ and $R_{21/10}(T)$ are plotted (blue and red, respectively).
The two sets are shifted slightly in the horizontal direction for clarity of the presentation.

The $R_{21/10}$ varies radially.
$R_{21/10}$ is high in the galactic center ($\sim 1$) and sharply declines toward the galactic radius of $R_{\rm gal}\sim$0.5~kpc (the edge of the center region).
It remains low ($\sim 0.6$) in $R_{\rm gal}\sim$0.5-2.0~kpc (the bar region).
It goes up higher ($\sim 0.7$) around $R_{\rm gal}\sim $2.0-3.0~kpc (at the bar end; boundary between the bar and spiral arm),
and continuously declines to a lower value ($\sim 0.6$) toward $R_{\rm gal}\gtrsim$6.0~kpc (the edge of the disk).
The definition of these radial ranges of the center, bar, disk are from \citet[][rounded]{Koda:2023aa}.
This radial trend is similar to the one observed in the Milky Way \citep{Sakamoto:1997aa}, as well as in some other barred spiral galaxies \citep{Maeda:2022aa, Egusa:2022aa, den-Brok:2023ab}.

\begin{figure}[h!]
\centering
\includegraphics[width=0.43\textwidth]{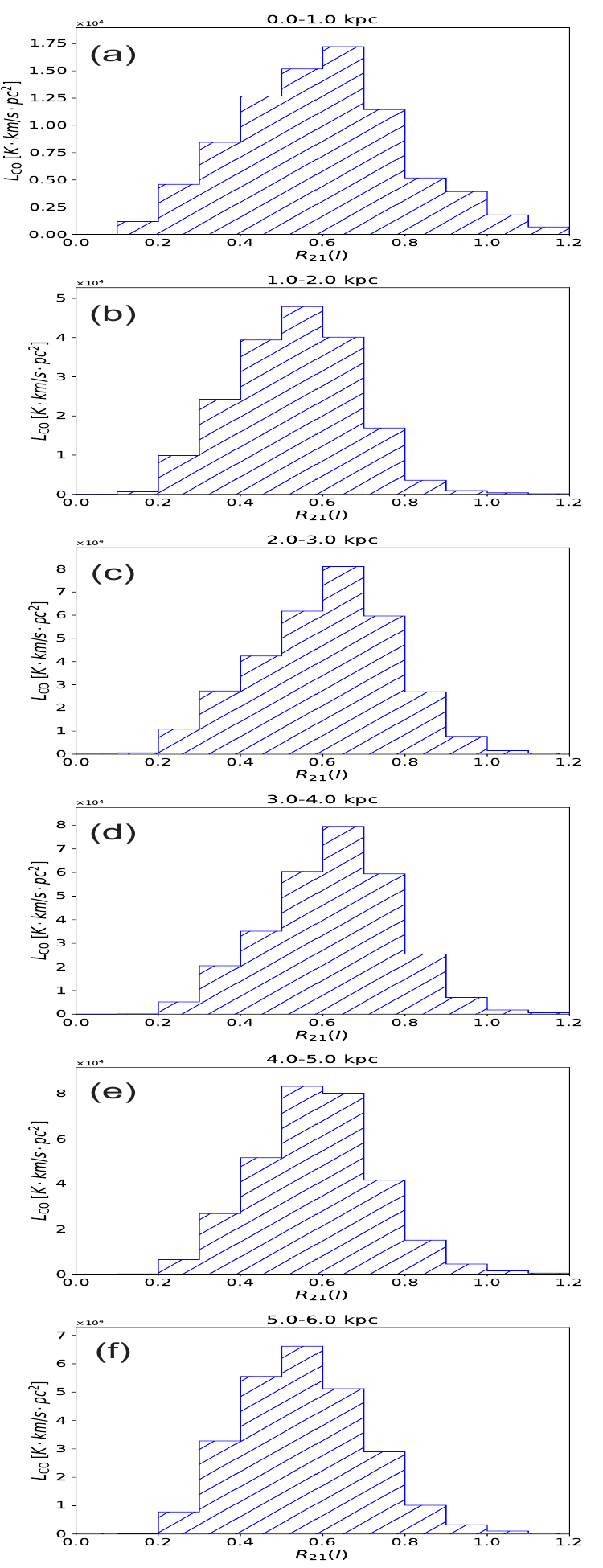}
\caption{
The same as Figure \ref{fig:R21hist_disk}a, but separated for each radial bin with 1~kpc width.
(a) galactic center,
(b) bar,
(c) bar end, and
(d)-(f) galactic disk where the spiral arms dominate.
The vertical axis is the CO(1-0) luminosity $L_{\rm CO}$ in each $R_{21/10}$ bin.
}
\label{fig:R21hist_bin}
\end{figure}

Figure \ref{fig:R21hist_bin} shows the histograms of $R_{21/10}$ in flux along $R_{\rm gal}$ in each 1~kpc radial bin.
The radial trend is again clear.
On average, $R_{21/10}$ is high in the center, becomes low in the bar, gets higher in the bar end, and slowly declines toward the disk edge.

The azimuthal variations of $R_{21/10}$ are more prominent than the radial variations from the bar to disk edge.
Indeed, Figure \ref{fig:R21rad}a,b show that the scatters at each radius are greater than the radial gradients.
Most parts of the azimuthal variations come from the systematic changes from the interarm regions to bar/spiral arms in the disk (Figure \ref{fig:R21map}).
In Figure \ref{fig:R21hist_bin}, $R_{21/10}$ takes the full range of about 0.3-1.0 in all $R_{\rm gal}$ (with some minor population outside this range),
and the fractional populations of high ($R_{21/10}>$0.7) and low ratio gas ($<0.7$) change along $R_{\rm gal}$.
Figure \ref{fig:R21hist_bin} also shows that the very high ratio gas ($R_{21/10}>1$) is rare except in the galactic center (panel a), as expected from the optical-depth consideration (Section \ref{sec:R21_1.0}).

The dominance of the azimuthal variations (e.g., arm/interarm variations) over the radial variations again suggests that the molecular gas evolves along the galactic rotation.
It starts from lower $n_{H_2}$ and $T_{\rm k}$ in the interarm regions and gets compressed to higher $n_{H_2}$ and $T_{\rm k}$, by about a factor of 2-3, in the bar/spiral arms.

\begin{deluxetable*}{ccccccccccccccccc}
\tablecaption{Statistics of $R_{\rm 21}$ in M83 \label{tab:R21stat}}
\tablewidth{0pt}
\tablehead{
\nocolhead{} & \multicolumn{7}{c}{$R_{21/10}(I)\equiv I_{21}/I_{10}$} & \nocolhead{} & \multicolumn{7}{c}{$R_{21/10}(T)\equiv T_{21}/T_{10}$} \\
\cline{2-8} \cline{10-16}
\nocolhead{} & \multicolumn{3}{c}{Intensity-weighted} & \nocolhead{} & \multicolumn{3}{c}{Area-weighted} & \nocolhead{} & \multicolumn{3}{c}{Intensity-weighted} & \nocolhead{} & \multicolumn{3}{c}{Area-weighted}  \\
\cline{2-4} \cline{6-8} \cline{10-12} \cline{14-16}
\colhead{$R_{\rm gal}$ [kpc]} & \colhead{Med} & \colhead{Ave} & \colhead{Std} & \nocolhead{} & \colhead{Med} & \colhead{Ave} & \colhead{Std} & \nocolhead{} & \colhead{Med} & \colhead{Ave} & \colhead{Std} & \nocolhead{} & \colhead{Med} & \colhead{Ave} & \colhead{Std}
}
\startdata
    All &  0.67 & 0.67 & 0.17 &&  0.59 & 0.59 & 0.16 &&  0.69 & 0.68 & 0.17 &&  0.62 & 0.61 & 0.19 \\
0.0-0.5 &  0.88 & 0.87 & 0.15 &&  0.76 & 0.75 & 0.21 &&  0.85 & 0.85 & 0.14 &&  0.75 & 0.74 & 0.21 \\
0.5-1.0 &  0.65 & 0.62 & 0.14 &&  0.54 & 0.54 & 0.16 &&  0.67 & 0.64 & 0.14 &&  0.59 & 0.56 & 0.18 \\
1.0-1.5 &  0.60 & 0.58 & 0.12 &&  0.56 & 0.55 & 0.15 &&  0.62 & 0.61 & 0.12 &&  0.60 & 0.58 & 0.16 \\
1.5-2.0 &  0.59 & 0.58 & 0.13 &&  0.52 & 0.52 & 0.14 &&  0.62 & 0.60 & 0.14 &&  0.57 & 0.55 & 0.18 \\
2.0-2.5 &  0.69 & 0.68 & 0.14 &&  0.61 & 0.60 & 0.16 &&  0.71 & 0.70 & 0.14 &&  0.65 & 0.62 & 0.19 \\
2.5-3.0 &  0.69 & 0.68 & 0.14 &&  0.64 & 0.62 & 0.17 &&  0.71 & 0.70 & 0.16 &&  0.67 & 0.65 & 0.21 \\
3.0-3.5 &  0.69 & 0.68 & 0.14 &&  0.65 & 0.64 & 0.16 &&  0.71 & 0.70 & 0.15 &&  0.68 & 0.65 & 0.19 \\
3.5-4.0 &  0.66 & 0.66 & 0.13 &&  0.62 & 0.62 & 0.15 &&  0.69 & 0.68 & 0.14 &&  0.66 & 0.64 & 0.18 \\
4.0-4.5 &  0.62 & 0.62 & 0.13 &&  0.59 & 0.59 & 0.15 &&  0.65 & 0.63 & 0.15 &&  0.62 & 0.61 & 0.18 \\
4.5-5.0 &  0.62 & 0.61 & 0.13 &&  0.58 & 0.58 & 0.15 &&  0.63 & 0.63 & 0.15 &&  0.61 & 0.59 & 0.19 \\
5.0-5.5 &  0.58 & 0.59 & 0.14 &&  0.55 & 0.55 & 0.15 &&  0.61 & 0.60 & 0.16 &&  0.57 & 0.56 & 0.18 \\
5.5-6.0 &  0.58 & 0.58 & 0.14 &&  0.56 & 0.56 & 0.16 &&  0.59 & 0.58 & 0.16 &&  0.57 & 0.56 & 0.18 \\
6.0-6.5 &  0.58 & 0.61 & 0.18 &&  0.56 & 0.58 & 0.19 &&  0.59 & 0.61 & 0.21 &&  0.57 & 0.58 & 0.23 \\ 
\hline
\enddata
\tablecomments{
Median (Med), average (Ave), and standard deviation (Std) are listed for the CO(1-0) intensity-weighted and area(pixel)-weighted calculations.
The left and right sections are for $R_{21/10}$ calculated with integrated intensities and peak surface brightness temperatures, respectively.
}
\end{deluxetable*}

\subsection{Relations to Star Formation and Galactic Structures} \label{sec:relation2SF}

\begin{figure*}[h!]
\centering
\includegraphics[width=1.0\textwidth]{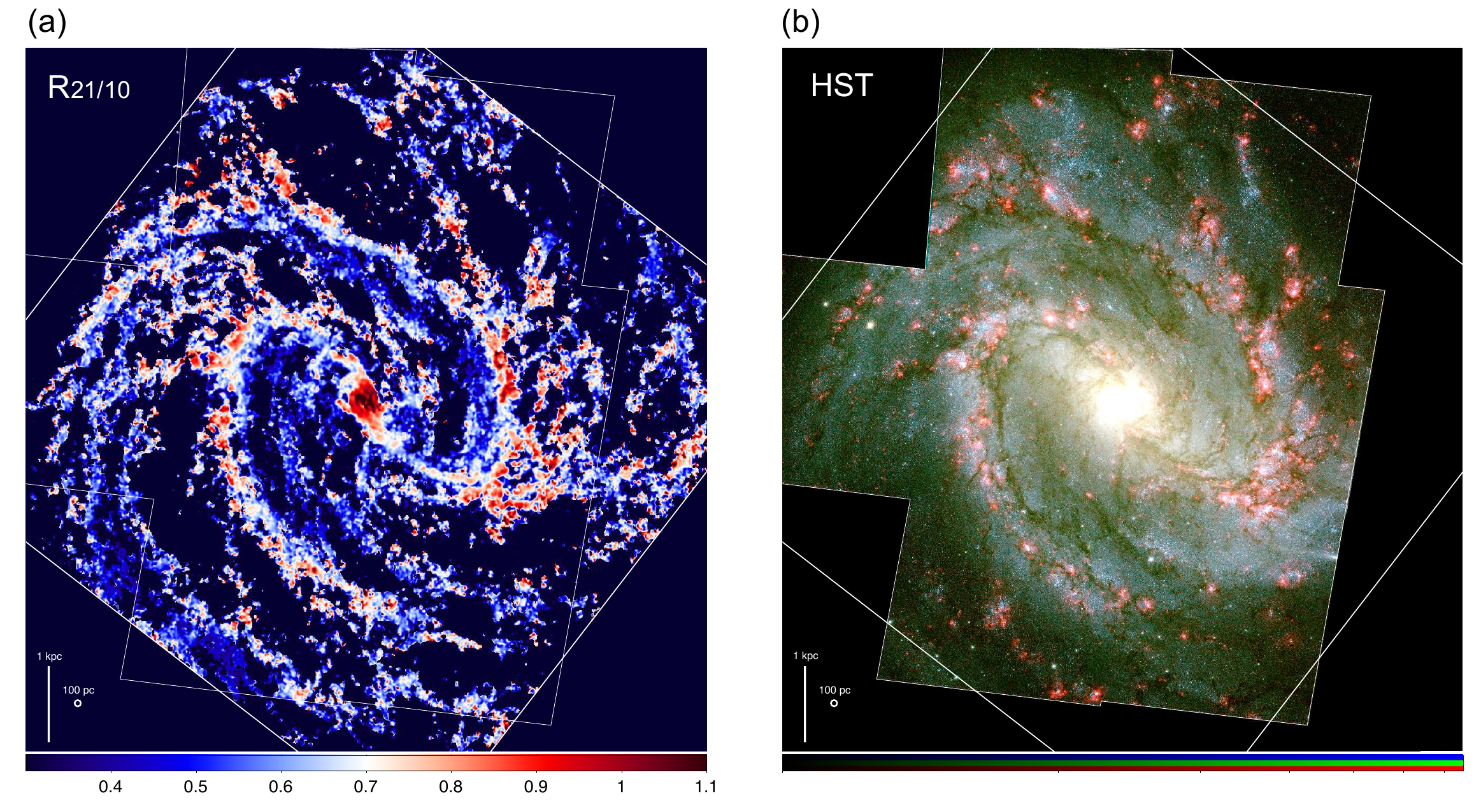}
\caption{
Comparison of $R_{21/10}(I)$ and Hubble Space Telescope (HST) image.
(a) The $R_{21/10}(I)$ map, the same as Figure \ref{fig:R21map}, but with the data coverages.
(b) The HST image with the F657N (H$\alpha$) image for R, a geometric mean of the F438W and F814W images for G, and the F438W image for B.
The white solid lines show the CO coverage (square shape) and HST coverage (irregular).}
\label{fig:hst}
\end{figure*}

The $R_{21/10}$ variations with respect to the galactic structures occur even without SF.
The sites of stellar feedback appear localized, and their impacts appear limited in their vicinity.
Figure \ref{fig:hst} compares the $R_{21/10}$ and optical image from the Hubble Space Telescope (HST) with red for the H$\alpha$ (F657N) filter \citep{Blair:2014aa}.
$R_{21/10}$ increases to $\gtrsim 0.7$ in the spiral arms, even in the portions without SF  (see the white and red regions in the left panel).
It is enhanced further to $R_{21/10}=$0.8-1.0 (red) when SF exists.
Indeed, the gas with the high ratio is almost entirely associated with HII regions and is localized in their vicinity.
From visual inspection, the radii of $R_{21/10}>0.8$ regions are $\lesssim$100~pc even around prominent HII regions.
They are predominantly on the convex side of the spiral arms (i.e., the downstream side assuming that the gas passes the spiral arms in the clockwise direction).
This suggests that the stellar feedback can influence molecular gas only in its vicinity and only when it is triggered.

\begin{figure*}[h!]
\centering
\includegraphics[width=0.9\textwidth]{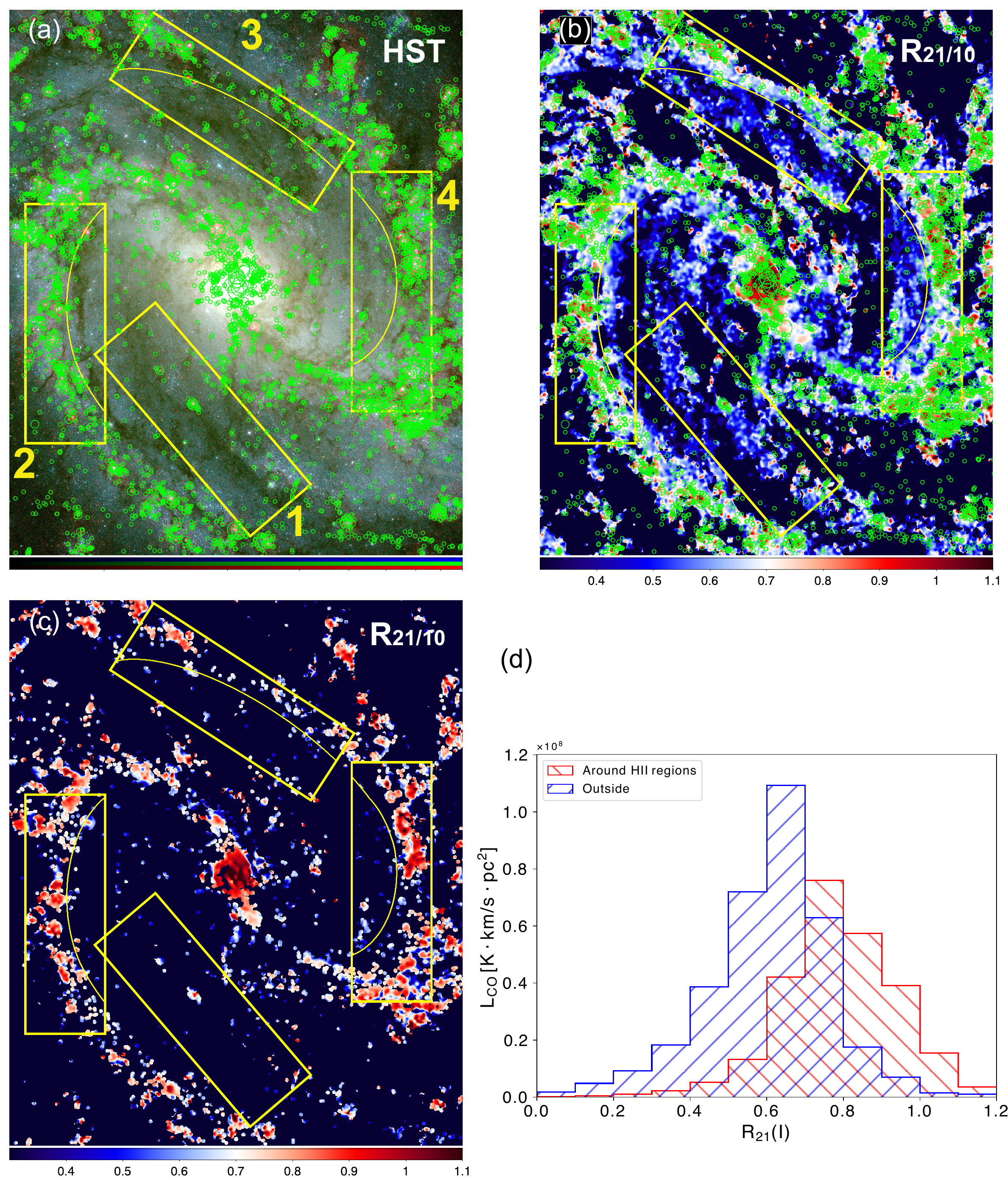}
\caption{
A central $4.35\arcmin \times 5.32\arcmin$ ($5.69\times 6.96\,\rm kpc^2$) region (a rectangular cutout that avoids the blanks in the HST images).
The HII regions are marked with green circles (``HII-region mask"), which separate the areas around and outside the HII regions (see Appendix \ref{sec:HIImask} for the mask generation).
The smallest circles have a diameter of 50~pc, and the resolution of the $R_{21/10}$ map is 46~pc.
Yellow boxes define Fields 1-4 ($3\times 1\,\rm kpc^2$) for discussions primarily in Section \ref{sec:noSFgas}.
(a) The HST image with H$\alpha$ in red (the same as Figure \ref{fig:hst}).
The green circles encircle or hide HII regions (red) and work as a mask of the HII regions.
(b) The $R_{21/10}$ map with the HII-region mask. $R_{21/10}$ reaches 0.7 (white) or higher on the spiral arms even outside the HII regions (see Fields 2, 3, and 4).
(c) The $R_{21/10}$ map, but only with the pixels within the HII-region mask.
$R_{21/10}$ is often elevated to $\sim$0.8-1.0.
(d) The histograms of $R_{21/10}$ in/around (red) and outside (blue) the HII regions (i.e., in and outside the HII-region mask, respectively).
}
\label{fig:HIImask}
\end{figure*}

To confirm these results further,
Figure \ref{fig:HIImask} separates the areas around and outside the HII regions (i.e., massive SF) detected in the HST image (see Appendix \ref{sec:HIImask}).
Figure \ref{fig:HIImask}a shows the HST image with H$\alpha$ in red.
The green circles are the detected HII regions (the HII-region mask), which nicely enclose or hide the H$\alpha$ emissions.
The detections reach as faint as an H$\alpha$ luminosity of $L_{\rm H\alpha}\sim 10^{35}\rm \, erg/s$.
As a reference, the Orion Nebula (M42/M43) is among the least impressive Galactic OB associations \citep{Hillenbrand:1997aa} and has $L_{\rm H\alpha} \sim 7\times 10^{36}\,\rm erg/s$ \citep{Scoville:2001aa}.
It would be detected even under an extinction of $A_{\rm V}\sim$5-6~mag.
The diameters of the green circles are adjusted with $L_{\rm H\alpha}$ to enclose the whole extents of the H$\alpha$ emissions: 50, 100, and 200~pc for $L_{\rm H\alpha}<10^{37}$, $10^{37-38}$, and $\geq 10^{38}\,\rm erg/s$, respectively \citep[][see their Figure 15]{Scoville:2001aa}.

Figure \ref{fig:HIImask}b shows the $R_{21/10}$ map with the HII-region mask (green circles).
Outside the mask, $R_{21/10}$ increases from the interarm regions into spiral arms even when they are away from the detected H$\alpha$ emissions.
Figure \ref{fig:HIImask}c isolates the regions within the mask. $R_{21/10}$ is generally high in/around the HII regions.
Figure \ref{fig:HIImask}d shows that $R_{21/10}$ can go up to $\sim 0.8$ outside the HII regions and is brought up to even a higher ratio of 0.8-1.0 around the HII regions.
Figure \ref{fig:HIImask} also boxes four example fields (yellow boxes; $3\times 1\kpc^2$ each): the interarm region (Field 1) and three regions around the spiral arms (Fields 2-4) [note these regions are defined also for discussion in Section \ref{sec:noSFgas}].
The total length of the spiral arms  in these regions is about 9~kpc.
The increase of $R_{21/10}$ from the interarm regions to the spiral arms is clearly seen.

\begin{figure*}[h!]
\centering
\includegraphics[width=1.0\textwidth]{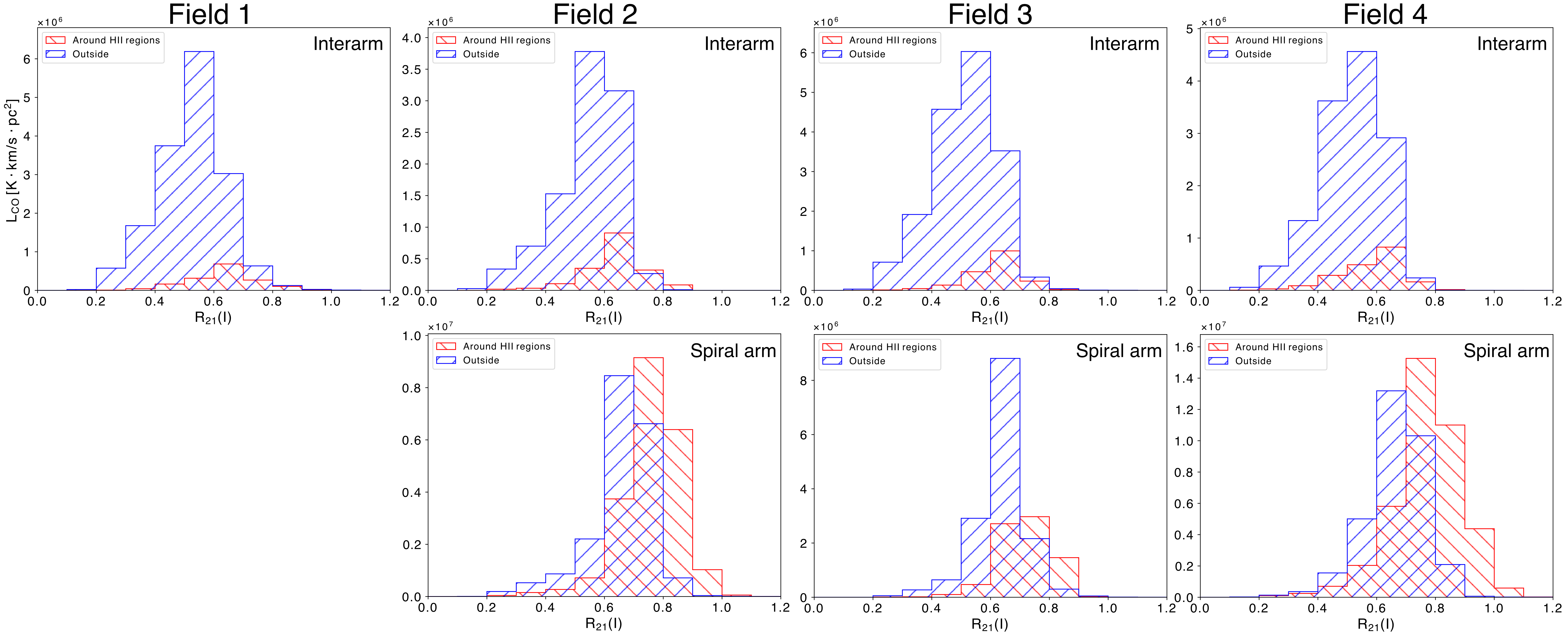}
\caption{
The same as Figure \ref{fig:HIImask}d, but for the interarm regions (top panels) and spiral arms (bottom) in Fields 1 (left panel) to 4 (right) defined in Figure \ref{fig:HIImask}.
Within each field, the regions in/around (red) and outside (blue) the HII regions (i.e., in and outside the HII-region mask, respectively) are separated.
}
\label{fig:HIImask_hist}
\end{figure*}

Albeit cursorily, Figure \ref{fig:HIImask} separates the spiral arms and interarm regions within Fields 2-4.
The curved lines within the boxes are the boundaries, whose convex sides are the spiral arm side.
Field 1 is considered as an interarm region entirely.
Figure \ref{fig:HIImask_hist} shows, in blue histograms, the increase of $R_{21/10}$ from the interarm regions (top panels) to spiral arms (bottom) even outside the HII regions, presumably suggesting the dynamical compression.
$R_{21/10}$ becomes even higher in/around the HII regions (red histograms), especially in the spiral arms (possibly due to more intense SF).
The fraction of molecular gas  in $L_{\rm CO}$ in/around the HII regions is as low as 9, 16, 10, and 13\% in the interarm regions of Fields 1-4, respectively.
It increases in the spiral arms in Fields 2-4, but is still only about half (52, 34, and 55\%). 
Therefore, the impact of the stellar feedback is limited.

We repeat that the spiral arm/interarm separation in the previous paragraph is cursory.
An accurate determination is beyond the scope of this work.
Our regions could be considered as the upstream and downstream sides of the gas flow across the spiral arms, instead of the interarm and spiral arm regions in an exact sense.
Even then, the above analysis illustrates that $R_{21/10}$ increases without SF as the molecular gas/clouds pass through the spiral arms, and that a large fraction of the molecular gas is not affected by SF.

\subsection{Relations with $I_{10}$ and $T_{10}$}

\begin{figure}[h!]
\centering
\includegraphics[width=0.48\textwidth]{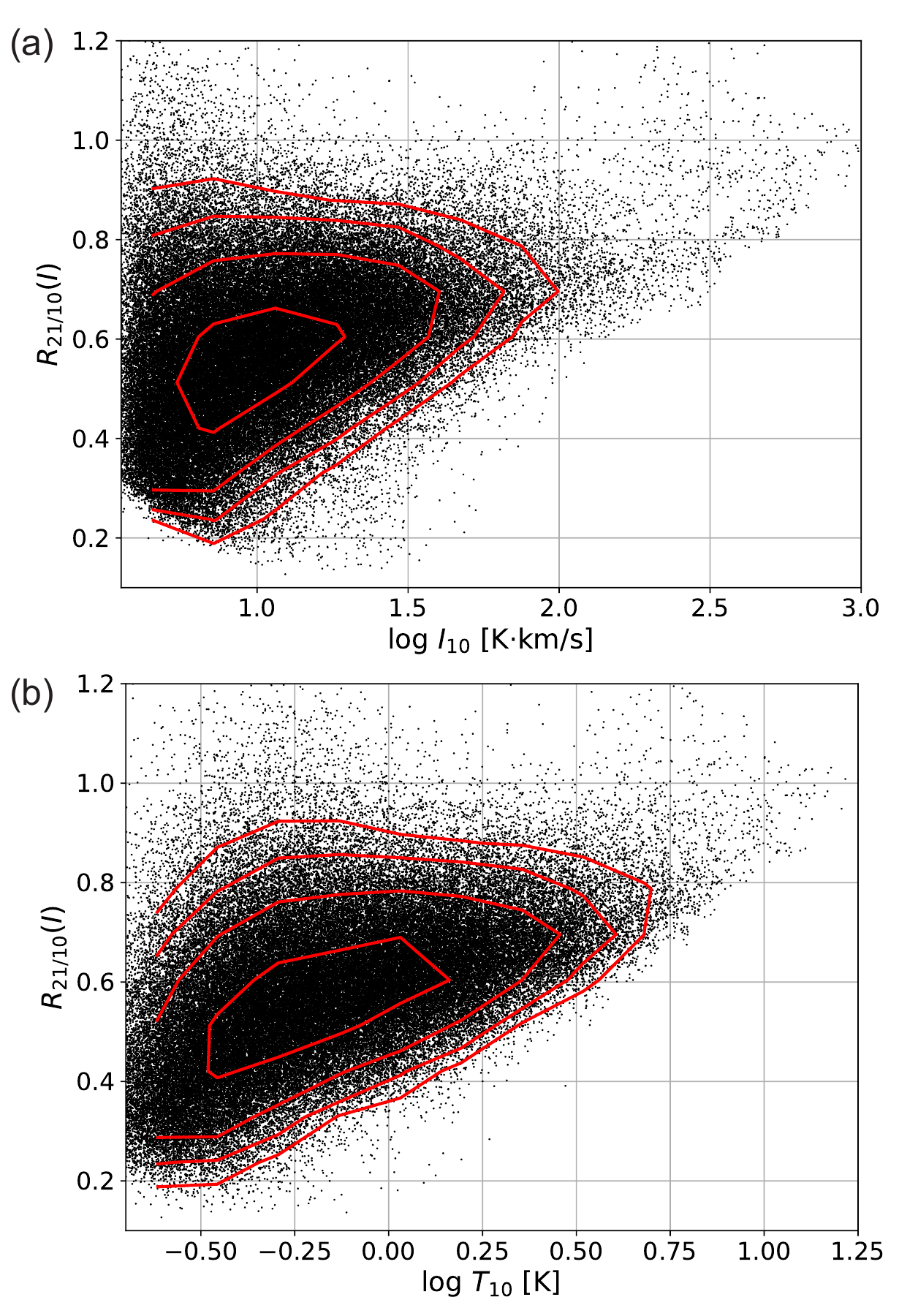}
\caption{$R_{21/10}(I)$ vs $I_{10}$ and $T_{10}$.
Only every 4th pixels (every 1") in Figures \ref{fig:co10co21moms} and \ref{fig:R21map} are plotted to reduce the density of the data points.
The red contours are drawn at 0.1, 0.2, 0.4, and 0.8 of the peak density.}
\label{fig:correlations}
\end{figure}

Figure \ref{fig:correlations} plots $R_{21/10}$ as a function of $I_{10}$ and $T_{10}$.
Again, only the 4th pixels in the RA and DEC directions are plotted to reduce the data density for clarity.
There is no straight-line correlation, but there are systematic trends.
$R_{21/10}$ tends to be skewed toward higher values at higher $I_{10}$ and $T_{10}$,
while it is spread over a larger range at lower $I_{10}$ and $T_{10}$, including high and low $R_{21/10}$.
The same trends were seen previously in the analyses of M51 and M83 at much lower resolutions \citep[$\sim$1~kpc; ][]{Koda:2012lr, Koda:2020aa} and in NGC 1300 with ALMA at a lower sensitivity \citep{Maeda:2022aa}.

The trends in Figure \ref{fig:correlations} also raise a caveat for a potential bias in $R_{21/10}$ analysis at lower sensitivity.
If the low intensity (or brightness) part in Figure \ref{fig:correlations} is cut by a sensitivity limit, the lower $R_{21/10}$ gas is selectively hidden, which biases the analysis toward higher $R_{21/10}$.
The relatively small (almost no) $R_{21/10}$ variations reported by \citet{den-Brok:2023ab} may be due to this bias.

\section{Discussion}

The large-scale $R_{21/10}$ pattern across the disk evidences that the physical conditions of the bulk molecular gas evolve as a response to galactic structures and dynamics.
The pattern is seen on a 46~pc resolution, close to the typical cloud diameter of 40~pc.
Thus, this evolution occurs within molecular clouds.
The bulk molecular gas within clouds systematically change its $n_{H_2}$ and/or $T_{\rm k}$ in the galactic structures and dynamics, by a factor of $\sim$2-3 around the typical values of ($n_{\rm H_2}$, $T_{\rm k}$)$\sim$($300{\,\rm cm^{-3}}$, $10{\,\rm K}$) between the interarm regions and bar/spiral arms.
This increase occurs without SF.
When HII regions exist, the $n_{H_2}$ and/or $T_{\rm k}$ appear to become even higher in the vicinity.

This contradicts the scenario of a very rapid molecular gas/cloud evolution \citep[e.g., ][]{Kruijssen:2019aa, Chevance:2020aa, Kim:2022aa},
as well as the strong emphasis on stellar feedback as the major driver of the molecular gas/cloud evolution \citep{Schinnerer:2024aa}.
The dynamical effects, seen in the $R_{21/10}$ map, take longer than the suggested short cloud lifetimes (Section \ref{sec:timescales}).
The impact of stellar feedback appears localized and limited in the galactic context (Section \ref{sec:hst}).
The presence of a large amount of dormant, non-star forming molecular gas also suggests the limited role of stellar feedback (Section \ref{sec:noSFgas}).
The galactic structures and dynamics appear to play more dominant roles than stellar feedback in the evolution of the bulk molecular gas (Section \ref{sec:dynamic}).

\citet{Koda:2009wd} discussed that, in galactic dynamics, molecular clouds can coagulate and/or fragment into larger or smaller molecular clouds.
They could be arranged in a sequence (i.e., chain of clouds) to form/become a filamentary structure.
In these dynamical processes, the gas remains largely molecular in the main parts of the disks in the MW and nearby galaxies \citep{Koda:2009wd, Koda:2016aa}.
This is an important difference between the dynamically-driven and feedback-driven pictures of molecular gas evolution.
We use the term ``cloud lifetime" for the full duration that the gas stays molecular in \textit{some} molecular clouds and remains in a state that it can potentially form stars,
even if the parent clouds coagulate and/or fragment \citep{Koda:2023aa}.

\subsection{Dynamical Timescales} \label{sec:timescales}

The large-scale $R_{21/10}$ pattern suggests that the molecular gas evolves in sync with galactic structures and dynamics,
along the flow from spiral arms (or bar) to interarm regions and then to the next spiral arms.
Therefore, in an order of magnitude, the evolution takes as long as the galactic rotation timescale, since most relevant dynamical processes operate in timescales close to the rotation timescale.

In the case of flat rotation (i.e., the angular speed is proportional to the galactic radius, $\Omega \propto R^{-1}$),
the galactic rotation timescale is
\begin{equation}
    t_{\rm rot} = \frac{2\pi}{\Omega}. \label{eq:trot}
\end{equation}
For M83, \citet{Koda:2023aa} estimated $t_{\rm rot} = 106{\,\rm Myr} (R/3{\,\rm kpc})(V/{174\kmps})^{-1}$
at $R=3{\,\rm kpc}$ with the observed rotation velocity of $V=174\kmps$.

The bar and spiral arms appear to be density-waves in the inner part of the disk \citep[$R<$3.5~kpc, ][]{Koda:2023aa}
as HII regions offset to their leading sides \citep{Egusa:2004aa, Egusa:2009aa}.
Hence, the rotation timescale (eq. \ref{eq:trot}) should be modified with respect to the bar/spiral pattern
\begin{equation}
      t_{\rm rot,p} = \frac{2\pi}{| \Omega-\Omega_{\rm p} |}, \label{eq:trotp}
\end{equation}
where the pattern speed $\Omega_{\rm p}$ reduces the denominator from eq. (\ref{eq:trot}) and makes $t_{\rm rot,p}\gg t_{\rm rot}$.
If we adopt $\Omega_{\rm p}=57\kmps\,\kpc^{-1}$ for M83 \citep{Hirota:2009vn}, we obtain $t_{\rm rot,p}=6$~Gyr at $R=3{\,\rm kpc}$.
It takes very long for the interarm molecular gas to reach the next spiral arm.

The spiral arms could be transient structures in the outer part of the disk \citep[$R>$3.5~kpc, ][]{Koda:2023aa}.
The transient spiral arms are driven by the swing amplification and by synchronizing the epicyclic motions \citep{Baba:2013vl, DOnghia:2013aa}.
They form and evolve in the timescale of epicyclic rotation,
\begin{equation}
    t_{\rm epi} \equiv \frac{2\pi}{\kappa} = 2\pi\left( R \frac{d\Omega^2}{dR} + 4\Omega^2 \right)^{-1/2} = \frac{\sqrt{2} \pi}{\Omega}, \label{eq:epi}
\end{equation}
where $\kappa$ is the epicyclic frequency.
It is $t_{\rm epi}=100{\,\rm Myr}(R/4{\,\rm kpc})(V/{174\kmps})^{-1}$ at a fiducial radius of $R=4{\,\rm kpc}$ for the outer disk region.

These dynamical timescales merely characterize how fast the gas (mass) can move, and hence, are the fundamental constraints required by the law of mass conservation (which any of more complex models must follow).
Therefore, the formation, maintenance, and destruction of the large-scale molecular structures (and the associated $R_{21/10}$ pattern) take as long as the rotation timescale in an order of magnitude.
If cloud lifetimes in these structures were to be as short as 5-30~Myr \citep[e.g., ][]{Kruijssen:2019aa, Chevance:2020aa, Kim:2022aa, Schinnerer:2024aa},
they would go through their life cycles within each of the long-lived molecular structures.
In that case, the $R_{21/10}$ map would not show the systematic, large-scale pattern.

\subsection{Limited Impacts of Stellar Feedback} \label{sec:hst}

The impact of stellar feedback appears localized and limited as discussed in Section \ref{sec:relation2SF}.
Figures \ref{fig:hst}, \ref{fig:HIImask}, and \ref{fig:HIImask_hist} show the increase of $R_{21/10}$ to $\gtrsim 0.7$ even in the portions of spiral arms without SF.
A higher ratio of $R_{21/10}=$0.8-1.0 is almost entirely associated with HII regions and is localized in their vicinity.
The radii of high $R_{21/10}>0.8$ regions are $\lesssim$100~pc even around prominent HII regions.
They are predominantly on the downstream side of the spiral arms assuming that the gas passes the arms in the clockwise direction.
Therefore, the stellar feedback influences the molecular gas only in its vicinity and only when it is triggered.

Although SF and high $R_{21/10}>$0.8-1.0 are clearly related, the causality remains unclear.
The high ratio could be caused by gas compression \textit{before} SF, or by stellar feedback \textit{after} SF.
In the former case, the high $R_{21/10}$ is not due to the feedback, and therefore, the sizes of the high $R_{21/10}$ regions are the upper limits of those of the regions impacted by stellar feedback.
This result casts an important caveat on the recent emphasis on stellar feedback as the dominant driver of the galactic molecular gas evolution \citep[][for reviews]{Chevance:2023aa, Schinnerer:2024aa}.

The above result does not mean that stellar feedback is unimportant.
Once SF occurs, its feedback could significantly affect the surrounding regions \citep[e.g., ][]{Chevance:2022aa, Deshmukh:2024aa} although the impact is only local.
In highly star-forming galaxies with prevalent SF, the impacts should also be prevalent \citep[e.g., ][]{Fisher:2019aa, Lenkic:2024aa}.
It is important to recognize that the bulk gas in normal nearby galaxies do not host prevalent (massive) SF.
Stellar feedback is important, but is not the main driver of the evolution of the bulk molecular gas across galaxies.

\subsection{Non-Star Forming Molecular Gas} \label{sec:noSFgas}

\citet{Koda:2023aa} showed the presence of large, massive, non-star forming molecular structures in the interarm regions of M83.
These structures often stretch over 1-4~kpc in length and contain $10^{7-8}\Msun$ in molecular gas mass, but have little/no SF.
They also have low $R_{21/10}$, and thus, have lower $n_{\rm H_2}$ and $T_{\rm k}$ than the average in the disk.
In fact, the low ratio gas ($<0.7$) spreads over a $3/4$ area of the full disk, which contains half the molecular gas mass (Section \ref{sec:R21overall}).
Figure \ref{fig:R21map} shows that the majority of this low $R_{21/10}$ gas resides in the interarm regions and on the upstream sides on/in the spiral arms assuming a clockwise disk rotation.

Figure \ref{fig:dormant} box four example fields of such massive dormant structures ($3\times1\kpc^2$ each; the same as in Figure \ref{fig:HIImask}), comparing
the CO(1-0) emission for the molecular gas,
$R_{21/10}$ for the physical conditions ($n_{\rm H_2}$ and $T_{\rm k}$),
optical and H$\alpha$ emission for dust extinction and exposed SF \citep[][see also Figures \ref{fig:HIImask} and \ref{fig:HIImask_hist}]{Blair:2014aa},
8$\mu$m emission for polycyclic aromatic hydrocarbons (PAHs),
24$\mu$m emission for heated dust/embedded SF \citep{Dale:2009aa}, and
HI 21cm emission for atomic hydrogen \citep{Lee:2024aa}.
Note that the HI gas emission is detected across the entire field except the galactic center where it is detected as absorption \citep{Lee:2024aa}.

Field 1 is part of the ``interarm molecular spiral arm" structure with a length of $\gtrsim 4$~kpc and mass of $\sim 10^8\Msun$.
Not many HII regions (exposed SF) nor heated dust emission (hidden SF) exist in the H$\alpha$ and 24$\mu$m images.
The brightest 24$\mu$m peak in this field serves as a gauge of low SF activity.
It is associated with the brightest HII region complex of $L_{\rm H\alpha}\sim 8\times 10^{37}{\,\rm erg/s}$ in this field (HST image),
which is already below the stochasticity limit where the SF rate becomes too low to fully populate the stellar initial mass function (IMF) \citep[$L_{\rm H\alpha}\sim 10^{38}$-$10^{39}{\,\rm erg/s}$; ][]{Kennicutt:2012aa}.
The rest of the field has SF activity below the 24$\mu$m peak.
Considering the IMF as a probability distribution function, massive stars, the source of the feedback, are least likely to form. 
This is consistent with the deficiency of HII regions in this region (Figure \ref{fig:HIImask}; detected down to the least active OB star formation even under $A_{\rm V}\sim$5-6~mag) and with the lower 24$\mu$m emission across the field.
$R_{21/10}$ is generally low in this massive molecular structure,
and therefore, the $n_{\rm H_2}$ and $T_{\rm k}$ are low across the 4~kpc length.
The atomic hydrogen (HI) is not particularly concentrated in this region, which also indicates that there has been little/no SF to dissociate the molecular gas.

\citet{Koda:2023aa} discussed that such large molecular structures (and underlying molecular gas/clouds) must be long-lived because of the long dynamical timescales.
It takes long ($\sim 100$~Myr) even only to sweep up the surrounding gas and to assemble their masses.
In addition, the large structures would not remain dormant ($R_{21/10}<0.7$) nor stay without SF if the molecular gas and clouds went through the rapid evolution driven by stellar feedback and lived shorter than the dynamical timescales.

Fields 2, 3, and 4 show the areas around the spiral arms.
They include the interarm regions on the upstream side of the spiral arms, assuming the clockwise gas flows.
Even before getting into the spiral arms, the gas is already molecular (the CO image in Figure \ref{fig:dormant}) without much enhancement in HI.
The molecular gas there shows lower $R_{21/10}$, and therefore, have lower $n_{\rm H_2}$ and $T_{\rm k}$ than the average over the disk.
The dust extinction in the HST image also suggests the presence of the gas.
The PAH 8$\mu$m emission is influenced by the background radiation field, but again shows the presence of the gas and dust.
However, there is little/no SF (H$\alpha$ and 24$\mu$m) in the molecular gas before it enters the spiral arms.

\subsection{The Dynamically-Driven Evolution} \label{sec:dynamic}

\begin{figure*}[h!]
\centering
\includegraphics[width=1.0\textwidth]{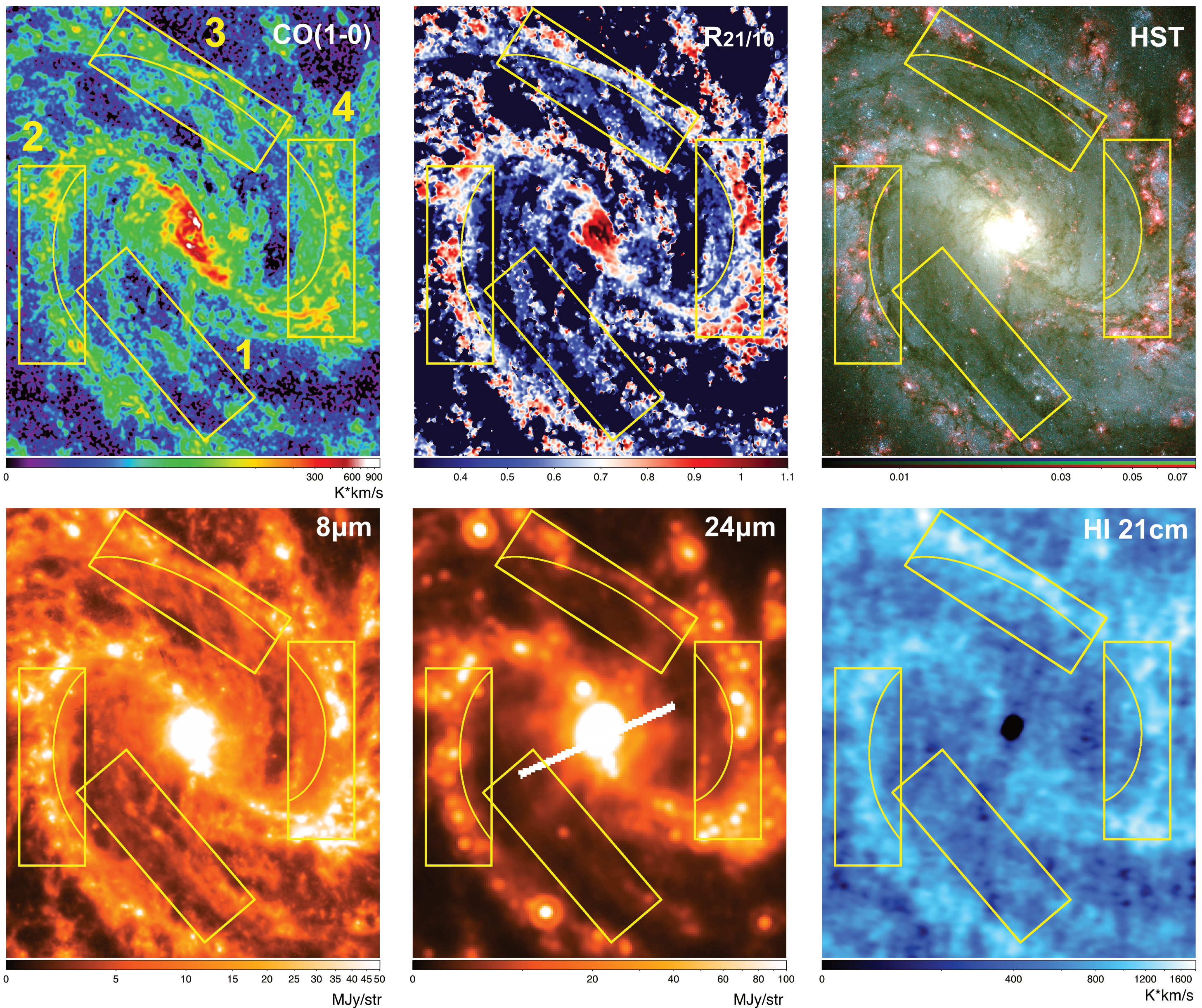}
\caption{
Multi-$\lambda$ comparisons for the area of Figure \ref{fig:HIImask}:
CO(1-0), $R_{21/10}$, optical image from HST with H$_{\alpha}$ emission in red \citep{Blair:2014aa}, 8$\mu m$ (PAH) and 24$\mu m$ (heated dust) emission from the Spitzer Space Telescope \citep{Dale:2009aa}, and HI 21cm emission from the Green Bank Telescope and Very Large Array \citep{Lee:2024aa}.
The four boxes are $3\times 1\kpc^2$ regions and contain massive, dormant, non-star forming molecular gas, as well as some star-forming molecular gas.
Field numbers are on the first panel.
Field 1 is the ``interarm molecular spiral arm" discussed in \citet{Koda:2023aa}.
Fields 2, 3, and 4 are segments of the spiral arms.
Assuming that the gas passes through the spiral arms in a clockwise direction, these fields include dormant, non-star forming molecular gas with low $R_{21/10}\lesssim 0.7$ in the upstream side of the spiral arms.
As the gas gets into the spiral arms, its $R_{21/10}\gtrsim 0.7$ increases even in the regions without SF.
Star-forming regions are localized in some parts along the spiral arms.
When SF exists, the associated $R_{21/10}$ appears even more elevated ($\sim$0.8-1.0).
See also Figures \ref{fig:HIImask} and \ref{fig:HIImask_hist}.}
\label{fig:dormant}
\end{figure*}

The findings of this paper point to the importance of galactic structures and dynamics in the evolution of molecular gas and SF.
The large-scale pattern in $R_{21/10}$ (Figures \ref{fig:R21map} and \ref{fig:hst}) suggests that the evolution of molecular gas is driven by galactic structures and dynamics.
$R_{21/10}$ is low ($<0.7$) before the spiral arms, and becomes higher ($\gtrsim 0.7$) as the molecular gas approaches/gets into the spiral arms (Section \ref{sec:R21}).
Thus, the $n_{\rm H_2}$ and $T_{\rm k}$ are low in the interarm regions but increase into the spiral arms.
This increase occurs even in the regions without much SF (most clearly in Field 3 in Figures \ref{fig:HIImask} and \ref{fig:dormant}).
Hence, it is not due to SF or feedback, but likely to the dynamical compression toward/in the spiral arms.

After this initial increase, some of the gas form massive stars and HII regions, and their surrounding $R_{21/10}$ is enhanced even higher to 0.8-1.0 (Figures \ref{fig:R21map}, \ref{fig:hst}, and \ref{fig:HIImask}).
$R_{21/10}$ could be enhanced either due to the dynamical compression \textit{before} SF, or gas heating by young stars \textit{after} SF \citep{Koda:2012lr}.

The presence of the dormant molecular gas in the interarm regions (Figures \ref{fig:HIImask} and \ref{fig:dormant}) suggests that the mere presence of molecular gas is not sufficient for SF.
The dynamically-driven evolution may result in the SF.

The large-scale pattern in $R_{21/10}$ also suggests that the molecular gas and clouds survive and travel across the disk.
If they were to go through the rapid evolution driven by stellar feedback and die immediately after their formation (i.e., short lifetimes), they would not move in the galactic disk and could not form the large-scale pattern in $R_{21/10}$.

\subsection{Similarity to Other Barred Spiral Galaxies} \label{sec:comp}

M83 is the closest morphological analog of the MW.
The $R_{21/10}$ variations in M83 are consistent with those found in the MW, although the previous Galactic studies were limited by the edge-on geometry as well as the data volume and quality \citep{Sakamoto:1994aa, Oka:1996aa, Sakamoto:1997aa, Hasegawa:1997lr, Oka:1998fk, Sawada:2001lr, Yoda:2010rf}.
The recent $R_{21/10}$ studies of nearby galaxies, even with ALMA, are also limited by low sensitivity and resolution.
However, with these limitations in mind, their $R_{21/10}$ maps show the large-scale variations similar to those in M83 in the barred spiral galaxies NGC 1300 \citep{Maeda:2022aa}, NGC 1365 \citep{Egusa:2022aa}, and NGC 3627 \citep{den-Brok:2023ab}.
They have high $R_{21/10}$ in the galactic centers, lower in the bar regions, and higher again at the bar ends.
These similarities indicate that the systematic variations in $R_{21/10}$ across galactic disks are common, at least among local barred spiral galaxies.

Most of these star-forming spiral galaxies have abundant gas, and the gas phase is typically molecular in their disks.
\citet{Koda:2016aa} pointed out that the evolution of molecular gas and clouds may depend on the environment, especially on the total amount of gas in the area (i.e., the average gas density in the environment) and gas phase (molecular vs atomic).
In fact, the lifetimes of molecular gas and clouds could be shorter in the low-density, atom-dominant environments, such as M33, the Large Magellanic Cloud, and the outskirts of the Milky Way \citep{Koda:2016aa}.
Clearly, we need a survey of a larger sample to investigate $R_{21/10}$ variations and molecular gas evolution.

\section{Summary and Conclusions}

We showed large-scale variations of the CO(2-1)/CO(1-0) line ratio ($R_{21/10}$) across the whole disk of the barred spiral galaxy M83 at a 46~pc resolution.
The large-scale pattern in $R_{21/10}$ is aligned with galactic structures, such as the bar, spiral arms, and interarm regions (e.g., Figure \ref{fig:R21map}).
The systematic variations evidence that the physical conditions of the bulk molecular gas evolve as a response to the galactic structures and dynamics, from the bar/spiral arms to interarm regions and then to the next bar/spiral arms.
This evolution occurs even in the regions without SF (e.g., Figure \ref{fig:HIImask_hist}).
Therefore, it is not primarily driven by SF nor feedback, but likely by the galactic structures and dynamics.
Some of the main findings are:

\begin{itemize}
\item On average, the molecular gas shows a typical ratio of around $R_{21/10}$=0.6-0.7, which roughly corresponds to the H$_2$ volume density and kinetic temperature of typical Galactic molecular clouds: ($n_{\rm H_2}$, $T_{\rm k}$) $\sim$ ($300\,\rm cm^{-3}$, $10\,\rm K$).
The large-scale variations across the disk occur around these values.
Generally, higher $n_{\rm H_2}$ and/or $T_{\rm k}$ result in higher $R_{21/10}$.

\item The low and high ratio gas ($R_{21/10}<0.7$ and $>0.7$) contains a similar amount of flux, and hence mass (58\% vs. 42 \%, respectively).
The low ratio gas spreads over a much larger area (76\% of the whole disk) compared to the high ratio gas (24\%).
The majority of the low $R_{21/10}$ gas resides in the interarm regions and on the upstream sides on/in the bar and spiral arms.

\item $R_{21/10}$ varies dominantly in the direction of galactic rotation.
It is lower ($<0.7$) in the interarm regions and is elevated in the bar ($\sim 0.7$) and spiral arms ($>0.7$),
which corresponds to a factor of $\sim$2-3 increases in $n_{\rm H_2}$ and/or $T_{\rm k}$ as the gas flows through the interarm regions into the bar and spiral arms.
These large-scale variations evidence that the gas physical conditions are evolving in sync with galactic rotation, structures, and dynamics.

\item Radial variations also exist, but are not as prominent except in the galactic center.
$R_{21/10}$ is high ($\sim$1.0) in the central $\sim$500~pc region, goes down ($\lesssim 0.7$) in the radial range of the bar, becomes high ($\gtrsim 0.7$) at the bar end, and gradually decreases  ($\lesssim 0.7$) toward the disk edge ($\sim 0.6$).

\item The resolution of 46~pc is close to the typical cloud diameter of 40~pc.
Therefore, the $R_{21/10}$ variations occur at the scale of molecular clouds.
They occur systematically over the populations of clouds in each galactic structure, again indicating the strong influence of large-scale galactic structures.

\item Recently, stellar feedback is emphasized as the major driver of molecular gas/cloud evolution \citep{Schinnerer:2024aa}.
However, its impact, as seen in elevated $R_{21/10}$=0.8-1.0, appears localized and limited.
The stellar feedback can influence the molecular gas only in its vicinity and only when it is triggered. It is not the controlling factor of the bulk molecular gas across galaxies.

\item Massive, large, non-star forming molecular structures exist, most notably in the interarm regions and on the upstream side of spiral arms.
They have low $R_{21/10}$, and hence, low $n_{\rm H_2}$ and/or $T_{\rm k}$.
They are not affected by stellar feedback.
It takes very long times ($\sim 100$~Myr) to assemble such structures, indicating their long lifetimes.

\item The large-scale pattern in $R_{21/10}$ also suggests that the molecular gas/clouds survive and travel over a substantial fraction of galactic rotation timescale ($>100$~Myr), as opposed to the recently-advocated rapid evolution scenario with short lifetimes \citep{Schinnerer:2024aa}.
Unless the molecular gas and clouds live for a long time, they cannot move and evolve in the galactic disk.

\item M83 is the closest morphological analog of the MW.
The $R_{21/10}$ variations in M83 are consistent with those found in the MW, although the previous MW studies were limited by the edge-on geometry as well as the data volume and quality.
Similar variations are also seen in other barred spiral galaxies.
These similarities indicate that the variations observed in M83 are common among local barred spiral galaxies.

\end{itemize}

\bibliographystyle{aasjournal}

\begin{thebibliography}{}
\expandafter\ifx\csname natexlab\endcsname\relax\def\natexlab#1{#1}\fi
\providecommand{\url}[1]{\href{#1}{#1}}
\providecommand{\dodoi}[1]{doi:~\href{http://doi.org/#1}{\nolinkurl{#1}}}
\providecommand{\doeprint}[1]{\href{http://ascl.net/#1}{\nolinkurl{http://ascl.net/#1}}}
\providecommand{\doarXiv}[1]{\href{https://arxiv.org/abs/#1}{\nolinkurl{https://arxiv.org/abs/#1}}}

\bibitem[{{Aalto} {et~al.}(1995){Aalto}, {Booth}, {Black}, \&
  {Johansson}}]{Aalto:1995lr}
{Aalto}, S., {Booth}, R.~S., {Black}, J.~H., \& {Johansson}, L.~E.~B. 1995,
  \aap, 300, 369

\bibitem[{{Baba} {et~al.}(2013){Baba}, {Saitoh}, \& {Wada}}]{Baba:2013vl}
{Baba}, J., {Saitoh}, T.~R., \& {Wada}, K. 2013, \apj, 763, 46,
  \dodoi{10.1088/0004-637X/763/1/46}

\bibitem[{{Bautista} {et~al.}(2023){Bautista}, {Koda}, {Yagi}, {Komiyama}, \&
  {Yamanoi}}]{Bautista:2023aa}
{Bautista}, J. M.~G., {Koda}, J., {Yagi}, M., {Komiyama}, Y., \& {Yamanoi}, H.
  2023, \apjs, 267, 10, \dodoi{10.3847/1538-4365/acd3e7}

\bibitem[{{Bertin} \& {Arnouts}(1996)}]{Bertin:1996lr}
{Bertin}, E., \& {Arnouts}, S. 1996, \aaps, 117, 393

\bibitem[{{Bigiel} {et~al.}(2008){Bigiel}, {Leroy}, {Walter}, {Brinks}, {de
  Blok}, {Madore}, \& {Thornley}}]{Bigiel:2008aa}
{Bigiel}, F., {Leroy}, A., {Walter}, F., {et~al.} 2008, \aj, 136, 2846,
  \dodoi{10.1088/0004-6256/136/6/2846}

\bibitem[{{Blair} {et~al.}(2014){Blair}, {Chandar}, {Dopita}, {Ghavamian},
  {Hammer}, {Kuntz}, {Long}, {Soria}, {Whitmore}, \& {Winkler}}]{Blair:2014aa}
{Blair}, W.~P., {Chandar}, R., {Dopita}, M.~A., {et~al.} 2014, \apj, 788, 55,
  \dodoi{10.1088/0004-637X/788/1/55}

\bibitem[{{Braine} \& {Combes}(1992)}]{Braine:1992lr}
{Braine}, J., \& {Combes}, F. 1992, \aap, 264, 433

\bibitem[{{Calamida} {et~al.}(2022){Calamida}, {Bajaj}, {Mack}, {Marinelli},
  {Medina}, {Pidgeon}, {Kozhurina-Platais}, {Shanahan}, \&
  {Som}}]{Calamida:2022aa}
{Calamida}, A., {Bajaj}, V., {Mack}, J., {et~al.} 2022, \aj, 164, 32,
  \dodoi{10.3847/1538-3881/ac73f0}

\bibitem[{{CASA Team} {et~al.}(2022){CASA Team}, {Bean}, {Bhatnagar}, {Castro},
  {Donovan Meyer}, {Emonts}, {Garcia}, {Garwood}, {Golap}, {Gonzalez Villalba},
  {Harris}, {Hayashi}, {Hoskins}, {Hsieh}, {Jagannathan}, {Kawasaki},
  {Keimpema}, {Kettenis}, {Lopez}, {Marvil}, {Masters}, {McNichols},
  {Mehringer}, {Miel}, {Moellenbrock}, {Montesino}, {Nakazato}, {Ott}, {Petry},
  {Pokorny}, {Raba}, {Rau}, {Schiebel}, {Schweighart}, {Sekhar}, {Shimada},
  {Small}, {Steeb}, {Sugimoto}, {Suoranta}, {Tsutsumi}, {van Bemmel},
  {Verkouter}, {Wells}, {Xiong}, {Szomoru}, {Griffith}, {Glendenning}, \&
  {Kern}}]{CASA-Team:2022aa}
{CASA Team}, {Bean}, B., {Bhatnagar}, S., {et~al.} 2022, \pasp, 134, 114501,
  \dodoi{10.1088/1538-3873/ac9642}

\bibitem[{{Chevance} {et~al.}(2023){Chevance}, {Krumholz}, {McLeod},
  {Ostriker}, {Rosolowsky}, \& {Sternberg}}]{Chevance:2023aa}
{Chevance}, M., {Krumholz}, M.~R., {McLeod}, A.~F., {et~al.} 2023, in
  Astronomical Society of the Pacific Conference Series, Vol. 534, Protostars
  and Planets VII, ed. S.~{Inutsuka}, Y.~{Aikawa}, T.~{Muto}, K.~{Tomida}, \&
  M.~{Tamura}, 1, \dodoi{10.48550/arXiv.2203.09570}

\bibitem[{{Chevance} {et~al.}(2020){Chevance}, {Kruijssen}, {Hygate},
  {Schruba}, {Longmore}, {Groves}, {Henshaw}, {Herrera}, {Hughes}, {Jeffreson},
  {Lang}, {Leroy}, {Meidt}, {Pety}, {Razza}, {Rosolowsky}, {Schinnerer},
  {Bigiel}, {Blanc}, {Emsellem}, {Faesi}, {Glover}, {Haydon}, {Ho}, {Kreckel},
  {Lee}, {Liu}, {Querejeta}, {Saito}, {Sun}, {Usero}, \&
  {Utomo}}]{Chevance:2020aa}
{Chevance}, M., {Kruijssen}, J.~M.~D., {Hygate}, A. P.~S., {et~al.} 2020,
  \mnras, 493, 2872, \dodoi{10.1093/mnras/stz3525}

\bibitem[{{Chevance} {et~al.}(2022){Chevance}, {Kruijssen}, {Krumholz},
  {Groves}, {Keller}, {Hughes}, {Glover}, {Henshaw}, {Herrera}, {Kim}, {Leroy},
  {Pety}, {Razza}, {Rosolowsky}, {Schinnerer}, {Schruba}, {Barnes}, {Bigiel},
  {Blanc}, {Dale}, {Emsellem}, {Faesi}, {Grasha}, {Klessen}, {Kreckel}, {Liu},
  {Longmore}, {Meidt}, {Querejeta}, {Saito}, {Sun}, \&
  {Usero}}]{Chevance:2022aa}
{Chevance}, M., {Kruijssen}, J.~M.~D., {Krumholz}, M.~R., {et~al.} 2022,
  \mnras, 509, 272, \dodoi{10.1093/mnras/stab2938}

\bibitem[{{Crosthwaite} \& {Turner}(2007)}]{Crosthwaite:2007uq}
{Crosthwaite}, L.~P., \& {Turner}, J.~L. 2007, \aj, 134, 1827,
  \dodoi{10.1086/521645}

\bibitem[{{Crosthwaite} {et~al.}(2002){Crosthwaite}, {Turner}, {Buchholz},
  {Ho}, \& {Martin}}]{Crosthwaite:2002yu}
{Crosthwaite}, L.~P., {Turner}, J.~L., {Buchholz}, L., {Ho}, P.~T.~P., \&
  {Martin}, R.~N. 2002, \aj, 123, 1892, \dodoi{10.1086/339479}

\bibitem[{{Dale} {et~al.}(2009){Dale}, {Cohen}, {Johnson}, {Schuster},
  {Calzetti}, {Engelbracht}, {Gil de Paz}, {Kennicutt}, {Lee}, {Begum},
  {Block}, {Dalcanton}, {Funes}, {Gordon}, {Johnson}, {Marble}, {Sakai},
  {Skillman}, {van Zee}, {Walter}, {Weisz}, {Williams}, {Wu}, \&
  {Wu}}]{Dale:2009aa}
{Dale}, D.~A., {Cohen}, S.~A., {Johnson}, L.~C., {et~al.} 2009, \apj, 703, 517,
  \dodoi{10.1088/0004-637X/703/1/517}

\bibitem[{{Dame} {et~al.}(2001){Dame}, {Hartmann}, \& {Thaddeus}}]{Dame:2001gs}
{Dame}, T.~M., {Hartmann}, D., \& {Thaddeus}, P. 2001, \apj, 547, 792,
  \dodoi{10.1086/318388}

\bibitem[{{Dame} {et~al.}(1987){Dame}, {Ungerechts}, {Cohen}, {de Geus},
  {Grenier}, {May}, {Murphy}, {Nyman}, \& {Thaddeus}}]{Dame:1987aa}
{Dame}, T.~M., {Ungerechts}, H., {Cohen}, R.~S., {et~al.} 1987, \apj, 322, 706,
  \dodoi{10.1086/165766}

\bibitem[{{den Brok} {et~al.}(2024){den Brok}, {Jim{\'e}nez-Donaire}, {Leroy},
  {Schinnerer}, {Bigiel}, {Pety}, {Petitpas}, {Usero}, {Teng}, {Humire},
  {Koch}, {Rosolowsky}, {Sandstrom}, {Liu}, {Zhang}, {Stuber}, {Chevance},
  {Dale}, {Eibensteiner}, {Gali{\'c}}, {Glover}, {Pan}, {Querejeta}, {Smith},
  {Williams}, {Wilner}, \& {Zhang}}]{den-Brok:2024aa}
{den Brok}, J., {Jim{\'e}nez-Donaire}, M.~J., {Leroy}, A., {et~al.} 2024, arXiv
  e-prints, arXiv:2410.21399, \dodoi{10.48550/arXiv.2410.21399}

\bibitem[{{den Brok} {et~al.}(2021){den Brok}, {Chatzigiannakis}, {Bigiel},
  {Puschnig}, {Barnes}, {Leroy}, {Jim{\'e}nez-Donaire}, {Usero}, {Schinnerer},
  {Rosolowsky}, {Faesi}, {Grasha}, {Hughes}, {Kruijssen}, {Liu}, {Neumann},
  {Pety}, {Querejeta}, {Saito}, {Schruba}, \& {Stuber}}]{den-Brok:2021aa}
{den Brok}, J.~S., {Chatzigiannakis}, D., {Bigiel}, F., {et~al.} 2021, \mnras,
  504, 3221, \dodoi{10.1093/mnras/stab859}

\bibitem[{{den Brok} {et~al.}(2023){den Brok}, {Leroy}, {Usero}, {Schinnerer},
  {Rosolowsky}, {Koch}, {Querejeta}, {Liu}, {Bigiel}, {Barnes}, {Chevance},
  {Colombo}, {Dale}, {Glover}, {Jimenez-Donaire}, {Teng}, \&
  {Williams}}]{den-Brok:2023ab}
{den Brok}, J.~S., {Leroy}, A.~K., {Usero}, A., {et~al.} 2023, \mnras,
  \dodoi{10.1093/mnras/stad3091}

\bibitem[{{Deshmukh} {et~al.}(2024){Deshmukh}, {Linden}, {Calzetti}, {Adamo},
  {Messa}, {Grasha}, {Sabbi}, {Smith}, \& {Johnson}}]{Deshmukh:2024aa}
{Deshmukh}, S., {Linden}, S.~T., {Calzetti}, D., {et~al.} 2024, \apjl, 974,
  L24, \dodoi{10.3847/2041-8213/ad7ba9}

\bibitem[{{D'Onghia} {et~al.}(2013){D'Onghia}, {Vogelsberger}, \&
  {Hernquist}}]{DOnghia:2013aa}
{D'Onghia}, E., {Vogelsberger}, M., \& {Hernquist}, L. 2013, \apj, 766, 34,
  \dodoi{10.1088/0004-637X/766/1/34}

\bibitem[{{Egusa} {et~al.}(2022){Egusa}, {Gao}, {Morokuma-Matsui}, {Liu}, \&
  {Maeda}}]{Egusa:2022aa}
{Egusa}, F., {Gao}, Y., {Morokuma-Matsui}, K., {Liu}, G., \& {Maeda}, F. 2022,
  \apj, 935, 64, \dodoi{10.3847/1538-4357/ac8050}

\bibitem[{{Egusa} {et~al.}(2009){Egusa}, {Kohno}, {Sofue}, {Nakanishi}, \&
  {Komugi}}]{Egusa:2009aa}
{Egusa}, F., {Kohno}, K., {Sofue}, Y., {Nakanishi}, H., \& {Komugi}, S. 2009,
  \apj, 697, 1870, \dodoi{10.1088/0004-637X/697/2/1870}

\bibitem[{{Egusa} {et~al.}(2004){Egusa}, {Sofue}, \&
  {Nakanishi}}]{Egusa:2004aa}
{Egusa}, F., {Sofue}, Y., \& {Nakanishi}, H. 2004, \pasj, 56, L45,
  \dodoi{10.1093/pasj/56.6.L45}

\bibitem[{{Fisher} {et~al.}(2019){Fisher}, {Bolatto}, {White}, {Glazebrook},
  {Abraham}, \& {Obreschkow}}]{Fisher:2019aa}
{Fisher}, D.~B., {Bolatto}, A.~D., {White}, H., {et~al.} 2019, \apj, 870, 46,
  \dodoi{10.3847/1538-4357/aaee8b}

\bibitem[{{Goldreich} \& {Kwan}(1974)}]{Goldreich:1974ab}
{Goldreich}, P., \& {Kwan}, J. 1974, \apj, 189, 441, \dodoi{10.1086/152821}

\bibitem[{{Goldsmith} {et~al.}(1983){Goldsmith}, {Young}, \&
  {Langer}}]{Goldsmith:1983aa}
{Goldsmith}, P.~F., {Young}, J.~S., \& {Langer}, W.~D. 1983, \apjs, 51, 203,
  \dodoi{10.1086/190845}

\bibitem[{{Hasegawa}(1997)}]{Hasegawa:1997lr}
{Hasegawa}, T. 1997, in IAU Symposium, Vol. 170, IAU Symposium, ed.
  {W.~B.~Latter, S.~J.~E.~Radford, P.~R.~Jewell, J.~G.~Mangum, \& J.~Bally},
  39--46

\bibitem[{{Hassani} {et~al.}(2023){Hassani}, {Rosolowsky}, {Leroy}, {Boquien},
  {Lee}, {Barnes}, {Belfiore}, {Bigiel}, {Cao}, {Chevance}, {Dale}, {Egorov},
  {Emsellem}, {Faesi}, {Grasha}, {Kim}, {Klessen}, {Kreckel}, {Kruijssen},
  {Larson}, {Meidt}, {Sandstrom}, {Schinnerer}, {Thilker}, {Watkins},
  {Whitmore}, \& {Williams}}]{Hassani:2023aa}
{Hassani}, H., {Rosolowsky}, E., {Leroy}, A.~K., {et~al.} 2023, \apjl, 944,
  L21, \dodoi{10.3847/2041-8213/aca8ab}

\bibitem[{{Hillenbrand}(1997)}]{Hillenbrand:1997aa}
{Hillenbrand}, L.~A. 1997, \aj, 113, 1733, \dodoi{10.1086/118389}

\bibitem[{{Hirota} {et~al.}(2009){Hirota}, {Kuno}, {Sato}, {Nakanishi},
  {Tosaki}, {Matsui}, {Habe}, \& {Sorai}}]{Hirota:2009vn}
{Hirota}, A., {Kuno}, N., {Sato}, N., {et~al.} 2009, \pasj, 61, 441,
  \dodoi{10.1093/pasj/61.3.441}

\bibitem[{{Hirota} {et~al.}(2024){Hirota}, {Koda}, {Egusa}, {Sawada},
  {Sakamoto}, {Heyer}, {Lee}, {Maeda}, {Boissier}, {Calzetti}, {Elmegreen},
  {Harada}, {Ho}, {Kobayashi}, {Kuno}, {Madore}, {Mart{\'\i}n}, {Donovan
  Meyer}, {Muraoka}, \& {Watanabe}}]{Hirota:2024aa}
{Hirota}, A., {Koda}, J., {Egusa}, F., {et~al.} 2024, arXiv e-prints,
  arXiv:2410.05424.
\newblock \doarXiv{2410.05424}

\bibitem[{{Keenan} {et~al.}(2024){Keenan}, {Marrone}, {Keating}, {Mayer},
  {Bays}, {Downey}, {Dunn}, {Flores}, {Folkers}, {Forbes}, {Guvenen},
  {Holmstedt}, {Moulton}, \& {Sullivan}}]{Keenan:2024aa}
{Keenan}, R.~P., {Marrone}, D.~P., {Keating}, G.~K., {et~al.} 2024, \apj, 975,
  150, \dodoi{10.3847/1538-4357/ad7504}

\bibitem[{{Kennicutt} \& {Evans}(2012)}]{Kennicutt:2012aa}
{Kennicutt}, R.~C., \& {Evans}, N.~J. 2012, \araa, 50, 531,
  \dodoi{10.1146/annurev-astro-081811-125610}

\bibitem[{{Kim} {et~al.}(2022){Kim}, {Chevance}, {Kruijssen}, {Leroy},
  {Schruba}, {Barnes}, {Bigiel}, {Blanc}, {Cao}, {Congiu}, {Dale}, {Faesi},
  {Glover}, {Grasha}, {Groves}, {Hughes}, {Klessen}, {Kreckel}, {McElroy},
  {Pan}, {Pety}, {Querejeta}, {Razza}, {Rosolowsky}, {Saito}, {Schinnerer},
  {Sun}, {Tomi{\v{c}}i{\'c}}, {Usero}, \& {Williams}}]{Kim:2022aa}
{Kim}, J., {Chevance}, M., {Kruijssen}, J.~M.~D., {et~al.} 2022, \mnras, 516,
  3006, \dodoi{10.1093/mnras/stac2339}

\bibitem[{{Koda} {et~al.}(2016){Koda}, {Scoville}, \& {Heyer}}]{Koda:2016aa}
{Koda}, J., {Scoville}, N., \& {Heyer}, M. 2016, \apj, 823, 76,
  \dodoi{10.3847/0004-637X/823/2/76}

\bibitem[{{Koda} {et~al.}(2019){Koda}, {Teuben}, {Sawada}, {Plunkett}, \&
  {Fomalont}}]{Koda:2019aa}
{Koda}, J., {Teuben}, P., {Sawada}, T., {Plunkett}, A., \& {Fomalont}, E. 2019,
  \pasp, 131, 054505, \dodoi{10.1088/1538-3873/ab047e}

\bibitem[{{Koda} {et~al.}(2009){Koda}, {Scoville}, {Sawada}, {La Vigne},
  {Vogel}, {Potts}, {Carpenter}, {Corder}, {Wright}, {White}, {Zauderer},
  {Patience}, {Sargent}, {Bock}, {Hawkins}, {Hodges}, {Kemball}, {Lamb},
  {Plambeck}, {Pound}, {Scott}, {Teuben}, \& {Woody}}]{Koda:2009wd}
{Koda}, J., {Scoville}, N., {Sawada}, T., {et~al.} 2009, \apjl, 700, L132,
  \dodoi{10.1088/0004-637X/700/2/L132}

\bibitem[{{Koda} {et~al.}(2012){Koda}, {Scoville}, {Hasegawa}, {Calzetti},
  {Donovan Meyer}, {Egusa}, {Kennicutt}, {Kuno}, {Louie}, {Momose}, {Sawada},
  {Sorai}, \& {Umei}}]{Koda:2012lr}
{Koda}, J., {Scoville}, N., {Hasegawa}, T., {et~al.} 2012, \apj, 761, 41,
  \dodoi{10.1088/0004-637X/761/1/41}

\bibitem[{{Koda} {et~al.}(2020){Koda}, {Sawada}, {Sakamoto}, {Hirota}, {Egusa},
  {Boissier}, {Calzetti}, {Meyer}, {Elmegreen}, {de Paz}, {Harada}, {Ho},
  {Kobayashi}, {Kuno}, {Mart{\'\i}n}, {Muraoka}, {Nakanishi}, {Scoville},
  {Seibert}, {Vlahakis}, \& {Watanabe}}]{Koda:2020aa}
{Koda}, J., {Sawada}, T., {Sakamoto}, K., {et~al.} 2020, \apjl, 890, L10,
  \dodoi{10.3847/2041-8213/ab70b7}

\bibitem[{{Koda} {et~al.}(2023){Koda}, {Hirota}, {Egusa}, {Sakamoto}, {Sawada},
  {Heyer}, {Baba}, {Boissier}, {Calzetti}, {Meyer}, {Elmegreen}, {de Paz},
  {Harada}, {Ho}, {Kobayashi}, {Kuno}, {Lee}, {Madore}, {Maeda}, {Mart{\'\i}n},
  {Muraoka}, {Nakanishi}, {Onodera}, {Pineda}, {Scoville}, \&
  {Watanabe}}]{Koda:2023aa}
{Koda}, J., {Hirota}, A., {Egusa}, F., {et~al.} 2023, \apj, 949, 108,
  \dodoi{10.3847/1538-4357/acc65e}

\bibitem[{{Kruijssen} {et~al.}(2019){Kruijssen}, {Schruba}, {Chevance},
  {Longmore}, {Hygate}, {Haydon}, {McLeod}, {Dalcanton}, {Tacconi}, \& {van
  Dishoeck}}]{Kruijssen:2019aa}
{Kruijssen}, J.~M.~D., {Schruba}, A., {Chevance}, M., {et~al.} 2019, \nat, 569,
  519, \dodoi{10.1038/s41586-019-1194-3}

\bibitem[{{Lee} {et~al.}(2024){Lee}, {Koda}, {Hirota}, {Egusa}, \&
  {Heyer}}]{Lee:2024aa}
{Lee}, A.~M., {Koda}, J., {Hirota}, A., {Egusa}, F., \& {Heyer}, M. 2024, \apj,
  968, 97, \dodoi{10.3847/1538-4357/ad40a0}

\bibitem[{{Lenki{\'c}} {et~al.}(2024){Lenki{\'c}}, {Fisher}, {Bolatto},
  {Teuben}, {Levy}, {Sun}, {Herrera-Camus}, {Glazebrook}, {Obreschkow}, \&
  {Abraham}}]{Lenkic:2024aa}
{Lenki{\'c}}, L., {Fisher}, D.~B., {Bolatto}, A.~D., {et~al.} 2024, \apj, 976,
  88, \dodoi{10.3847/1538-4357/ad758c}

\bibitem[{{Leroy} {et~al.}(2008){Leroy}, {Walter}, {Brinks}, {Bigiel}, {de
  Blok}, {Madore}, \& {Thornley}}]{Leroy:2008fj}
{Leroy}, A.~K., {Walter}, F., {Brinks}, E., {et~al.} 2008, \aj, 136, 2782,
  \dodoi{10.1088/0004-6256/136/6/2782}

\bibitem[{{Leroy} {et~al.}(2013){Leroy}, {Walter}, {Sandstrom}, {Schruba},
  {Munoz-Mateos}, {Bigiel}, {Bolatto}, {Brinks}, {de Blok}, {Meidt}, {Rix},
  {Rosolowsky}, {Schinnerer}, {Schuster}, \& {Usero}}]{Leroy:2013aa}
{Leroy}, A.~K., {Walter}, F., {Sandstrom}, K., {et~al.} 2013, \aj, 146, 19,
  \dodoi{10.1088/0004-6256/146/2/19}

\bibitem[{{Leroy} {et~al.}(2021){Leroy}, {Schinnerer}, {Hughes}, {Rosolowsky},
  {Pety}, {Schruba}, {Usero}, {Blanc}, {Chevance}, {Emsellem}, {Faesi},
  {Herrera}, {Liu}, {Meidt}, {Querejeta}, {Saito}, {Sandstrom}, {Sun},
  {Williams}, {Anand}, {Barnes}, {Behrens}, {Belfiore}, {Benincasa},
  {Be{\v{s}}li{\'c}}, {Bigiel}, {Bolatto}, {den Brok}, {Cao}, {Chandar},
  {Chastenet}, {Chiang}, {Congiu}, {Dale}, {Deger}, {Eibensteiner}, {Egorov},
  {Garc{\'\i}a-Rodr{\'\i}guez}, {Glover}, {Grasha}, {Henshaw}, {Ho}, {Kepley},
  {Kim}, {Klessen}, {Kreckel}, {Koch}, {Kruijssen}, {Larson}, {Lee}, {Lopez},
  {Machado}, {Mayker}, {McElroy}, {Murphy}, {Ostriker}, {Pan}, {Pessa},
  {Puschnig}, {Razza}, {S{\'a}nchez-Bl{\'a}zquez}, {Santoro}, {Sardone},
  {Scheuermann}, {Sliwa}, {Sormani}, {Stuber}, {Thilker}, {Turner}, {Utomo},
  {Watkins}, \& {Whitmore}}]{Leroy:2021ab}
{Leroy}, A.~K., {Schinnerer}, E., {Hughes}, A., {et~al.} 2021, \apjs, 257, 43,
  \dodoi{10.3847/1538-4365/ac17f3}

\bibitem[{{Leroy} {et~al.}(2022){Leroy}, {Rosolowsky}, {Usero}, {Sandstrom},
  {Schinnerer}, {Schruba}, {Bolatto}, {Sun}, {Barnes}, {Belfiore}, {Bigiel},
  {den Brok}, {Cao}, {Chiang}, {Chevance}, {Dale}, {Eibensteiner}, {Faesi},
  {Glover}, {Hughes}, {Jim{\'e}nez Donaire}, {Klessen}, {Koch}, {Kruijssen},
  {Liu}, {Meidt}, {Pan}, {Pety}, {Puschnig}, {Querejeta}, {Saito}, {Sardone},
  {Watkins}, {Weiss}, \& {Williams}}]{Leroy:2022ac}
{Leroy}, A.~K., {Rosolowsky}, E., {Usero}, A., {et~al.} 2022, \apj, 927, 149,
  \dodoi{10.3847/1538-4357/ac3490}

\bibitem[{{Long} {et~al.}(2022){Long}, {Blair}, {Winkler}, {Della Bruna},
  {Adamo}, {McLeod}, \& {Amram}}]{Long:2022aa}
{Long}, K.~S., {Blair}, W.~P., {Winkler}, P.~F., {et~al.} 2022, \apj, 929, 144,
  \dodoi{10.3847/1538-4357/ac5aa3}

\bibitem[{{Lundgren} {et~al.}(2004){Lundgren}, {Wiklind}, {Olofsson}, \&
  {Rydbeck}}]{Lundgren:2004aa}
{Lundgren}, A.~A., {Wiklind}, T., {Olofsson}, H., \& {Rydbeck}, G. 2004, \aap,
  413, 505, \dodoi{10.1051/0004-6361:20031507}

\bibitem[{{Maeda} {et~al.}(2022){Maeda}, {Egusa}, {Ohta}, {Fujimoto}, {Habe},
  \& {Asada}}]{Maeda:2022aa}
{Maeda}, F., {Egusa}, F., {Ohta}, K., {et~al.} 2022, \apj, 926, 96,
  \dodoi{10.3847/1538-4357/ac4505}

\bibitem[{{Nakanishi} \& {Sofue}(2016)}]{Nakanishi:2016aa}
{Nakanishi}, H., \& {Sofue}, Y. 2016, \pasj, 68, 5, \dodoi{10.1093/pasj/psv108}

\bibitem[{{Oka} {et~al.}(1996{\natexlab{a}}){Oka}, {Hasegawa}, {Handa},
  {Hayashi}, \& {Sakamoto}}]{Oka:1996lr}
{Oka}, T., {Hasegawa}, T., {Handa}, T., {Hayashi}, M., \& {Sakamoto}, S.
  1996{\natexlab{a}}, \apj, 460, 334, \dodoi{10.1086/176973}

\bibitem[{{Oka} {et~al.}(1996{\natexlab{b}}){Oka}, {Hasegawa}, {Handa},
  {Hayashi}, \& {Sakamoto}}]{Oka:1996aa}
---. 1996{\natexlab{b}}, \apj, 460, 334, \dodoi{10.1086/176973}

\bibitem[{{Oka} {et~al.}(1998){Oka}, {Hasegawa}, {Hayashi}, {Handa}, \&
  {Sakamoto}}]{Oka:1998fk}
{Oka}, T., {Hasegawa}, T., {Hayashi}, M., {Handa}, T., \& {Sakamoto}, S. 1998,
  \apj, 493, 730, \dodoi{10.1086/305133}

\bibitem[{{Saintonge} {et~al.}(2017){Saintonge}, {Catinella}, {Tacconi},
  {Kauffmann}, {Genzel}, {Cortese}, {Dav{\'e}}, {Fletcher},
  {Graci{\'a}-Carpio}, {Kramer}, {Heckman}, {Janowiecki}, {Lutz}, {Rosario},
  {Schiminovich}, {Schuster}, {Wang}, {Wuyts}, {Borthakur}, {Lamperti}, \&
  {Roberts-Borsani}}]{Saintonge:2017aa}
{Saintonge}, A., {Catinella}, B., {Tacconi}, L.~J., {et~al.} 2017, \apjs, 233,
  22, \dodoi{10.3847/1538-4365/aa97e0}

\bibitem[{{Sakamoto} {et~al.}(1997{\natexlab{a}}){Sakamoto}, {Hasegawa},
  {Handa}, {Hayashi}, \& {Oka}}]{Sakamoto:1997aa}
{Sakamoto}, S., {Hasegawa}, T., {Handa}, T., {Hayashi}, M., \& {Oka}, T.
  1997{\natexlab{a}}, \apj, 486, 276, \dodoi{10.1086/304479}

\bibitem[{{Sakamoto} {et~al.}(1997{\natexlab{b}}){Sakamoto}, {Hasegawa},
  {Hayashi}, {Morino}, \& {Sato}}]{Sakamoto:1997fk}
{Sakamoto}, S., {Hasegawa}, T., {Hayashi}, M., {Morino}, J.-I., \& {Sato}, K.
  1997{\natexlab{b}}, \apj, 481, 302, \dodoi{10.1086/304023}

\bibitem[{{Sakamoto} {et~al.}(1994){Sakamoto}, {Hayashi}, {Hasegawa}, {Handa},
  \& {Oka}}]{Sakamoto:1994aa}
{Sakamoto}, S., {Hayashi}, M., {Hasegawa}, T., {Handa}, T., \& {Oka}, T. 1994,
  \apj, 425, 641, \dodoi{10.1086/174011}

\bibitem[{{Sault} {et~al.}(1996){Sault}, {Staveley-Smith}, \&
  {Brouw}}]{Sault:1996uq}
{Sault}, R.~J., {Staveley-Smith}, L., \& {Brouw}, W.~N. 1996, \aaps, 120, 375

\bibitem[{{Sault} {et~al.}(1995){Sault}, {Teuben}, \& {Wright}}]{Sault:1995kl}
{Sault}, R.~J., {Teuben}, P.~J., \& {Wright}, M.~C.~H. 1995, in Astronomical
  Society of the Pacific Conference Series, Vol.~77, Astronomical Data Analysis
  Software and Systems IV, ed. {R.~A.~Shaw, H.~E.~Payne, \& J.~J.~E.~Hayes},
  433

\bibitem[{{Sawada} {et~al.}(2001){Sawada}, {Hasegawa}, {Handa}, {Morino},
  {Oka}, {Booth}, {Bronfman}, {Hayashi}, {Luna Castellanos}, {Nyman},
  {Sakamoto}, {Seta}, {Shaver}, {Sorai}, \& {Usuda}}]{Sawada:2001lr}
{Sawada}, T., {Hasegawa}, T., {Handa}, T., {et~al.} 2001, \apjs, 136, 189,
  \dodoi{10.1086/321793}

\bibitem[{{Schinnerer} \& {Leroy}(2024)}]{Schinnerer:2024aa}
{Schinnerer}, E., \& {Leroy}, A.~K. 2024, arXiv e-prints, arXiv:2403.19843,
  \dodoi{10.48550/arXiv.2403.19843}

\bibitem[{{Scoville} \& {Hersh}(1979)}]{Scoville:1979lg}
{Scoville}, N.~Z., \& {Hersh}, K. 1979, \apj, 229, 578, \dodoi{10.1086/156991}

\bibitem[{{Scoville} {et~al.}(2001){Scoville}, {Polletta}, {Ewald}, {Stolovy},
  {Thompson}, \& {Rieke}}]{Scoville:2001aa}
{Scoville}, N.~Z., {Polletta}, M., {Ewald}, S., {et~al.} 2001, \aj, 122, 3017,
  \dodoi{10.1086/323445}

\bibitem[{{Scoville} \& {Sanders}(1987)}]{Scoville:1987vo}
{Scoville}, N.~Z., \& {Sanders}, D.~B. 1987, in Astrophysics and Space Science
  Library, Vol. 134, Interstellar Processes, ed. {D.~J.~Hollenbach \&
  H.~A.~Thronson Jr.}, 21--50

\bibitem[{{Scoville} \& {Solomon}(1974)}]{Scoville:1974yu}
{Scoville}, N.~Z., \& {Solomon}, P.~M. 1974, \apjl, 187, L67,
  \dodoi{10.1086/181398}

\bibitem[{{Seta} {et~al.}(1998){Seta}, {Hasegawa}, {Dame}, {Sakamoto}, {Oka},
  {Handa}, {Hayashi}, {Morino}, {Sorai}, \& {Usuda}}]{Seta:1998aa}
{Seta}, M., {Hasegawa}, T., {Dame}, T.~M., {et~al.} 1998, \apj, 505, 286,
  \dodoi{10.1086/306141}

\bibitem[{{Sun} {et~al.}(2018){Sun}, {Leroy}, {Schruba}, {Rosolowsky},
  {Hughes}, {Kruijssen}, {Meidt}, {Schinnerer}, {Blanc}, {Bigiel}, {Bolatto},
  {Chevance}, {Groves}, {Herrera}, {Hygate}, {Pety}, {Querejeta}, {Usero}, \&
  {Utomo}}]{Sun:2018aa}
{Sun}, J., {Leroy}, A.~K., {Schruba}, A., {et~al.} 2018, \apj, 860, 172,
  \dodoi{10.3847/1538-4357/aac326}

\bibitem[{{Sun} {et~al.}(2020){Sun}, {Leroy}, {Schinnerer}, {Hughes},
  {Rosolowsky}, {Querejeta}, {Schruba}, {Liu}, {Saito}, {Herrera}, {Faesi},
  {Usero}, {Pety}, {Kruijssen}, {Ostriker}, {Bigiel}, {Blanc}, {Bolatto},
  {Boquien}, {Chevance}, {Dale}, {Deger}, {Emsellem}, {Glover}, {Grasha},
  {Groves}, {Henshaw}, {Jimenez-Donaire}, {Kim}, {Klessen}, {Kreckel}, {Lee},
  {Meidt}, {Sandstrom}, {Sardone}, {Utomo}, \& {Williams}}]{Sun:2020tt}
{Sun}, J., {Leroy}, A.~K., {Schinnerer}, E., {et~al.} 2020, \apjl, 901, L8,
  \dodoi{10.3847/2041-8213/abb3be}

\bibitem[{{van der Tak} {et~al.}(2007){van der Tak}, {Black}, {Sch{\"o}ier},
  {Jansen}, \& {van Dishoeck}}]{van-der-Tak:2007qy}
{van der Tak}, F.~F.~S., {Black}, J.~H., {Sch{\"o}ier}, F.~L., {Jansen}, D.~J.,
  \& {van Dishoeck}, E.~F. 2007, \aap, 468, 627,
  \dodoi{10.1051/0004-6361:20066820}

\bibitem[{{Vlahakis} {et~al.}(2013){Vlahakis}, {van der Werf}, {Israel}, \&
  {Tilanus}}]{Vlahakis:2013aa}
{Vlahakis}, C., {van der Werf}, P., {Israel}, F.~P., \& {Tilanus}, R.~P.~J.
  2013, \mnras, 433, 1837, \dodoi{10.1093/mnras/stt841}

\bibitem[{{Vogel} {et~al.}(1984){Vogel}, {Wright}, {Plambeck}, \&
  {Welch}}]{Vogel:1984aa}
{Vogel}, S.~N., {Wright}, M.~C.~H., {Plambeck}, R.~L., \& {Welch}, W.~J. 1984,
  \apj, 283, 655, \dodoi{10.1086/162351}

\bibitem[{{Yajima} {et~al.}(2021){Yajima}, {Sorai}, {Miyamoto}, {Muraoka},
  {Kuno}, {Kaneko}, {Takeuchi}, {Yasuda}, {Tanaka}, {Morokuma-Matsui}, \&
  {Kobayashi}}]{Yajima:2021aa}
{Yajima}, Y., {Sorai}, K., {Miyamoto}, Y., {et~al.} 2021, \pasj, 73, 257,
  \dodoi{10.1093/pasj/psaa119}

\bibitem[{{Yoda} {et~al.}(2010){Yoda}, {Handa}, {Kohno}, {Nakajima}, {Kaiden},
  {Yonekura}, {Ogawa}, {Morino}, \& {Dobashi}}]{Yoda:2010rf}
{Yoda}, T., {Handa}, T., {Kohno}, K., {et~al.} 2010, \pasj, 62, 1277,
  \dodoi{10.1093/pasj/62.5.1277}

\end{thebibliography}

\clearpage

\appendix

\section{Data Reduction and Imaging} \label{sec:reductiondetails}

The method of data reduction is discussed in depth by \citet{Koda:2023aa}.
We have improved the imaging part.
For both CO(1-0) and CO(2-1), the visibilities are calibrated as in \citet{Koda:2023aa}.
The TP cubes are reduced by \citet{Koda:2020aa} and are converted to visibilities using the Total Power to Visibilities (TP2VIS) package \citep{Koda:2019aa} at the mosaic pointing positions of the 12m array observations.
The joint-imaging of the 12m, 7m, and TP array data is performed with the Multichannel Image Reconstruction, Image Analysis, and Display package \citep[MIRIAD; ][]{Sault:1995kl, Sault:1996uq}.
The CLEAN algorithm in MIRIAD takes into account the spatially-variable point spread function (PSF), which is important for large-scale mosaic data.
The CO(1-0) observations employ the same set of 435 pointing positions for the 12m and 7m arrays.
The CO(2-1) observations use 1335 and 512 pointing positions with the 12m and 7m arrays, respectively.
All of these pointings are imaged together for each CO transition.
In this section, we discuss the parts of data reduction that are new and different from the method in \citet{Koda:2023aa}.

\subsection{CO(1-0)} \label{sec:CO10reduction}

The CO(1-0) data cube by \citet{Koda:2023aa} had remaining shallow negative sidebloes in some interarm regions.
We eliminate most of them by improving the imaging process (i.e., deeper CLEAN).
Such sidelobes are common among radio interferometry in general, and most of the published ALMA data of nearby galaxies in the literature suffer from such negative sidelobes at low levels.
Most of these emissions are artifacts and are the yet-CLEANed residuals ($\lesssim 1\sigma$) of more significant CLEANed emissions (like ``pedestal" of the stronger emissions).
They are extended over large areas and carry significant total fluxes and generate the negative sidelobes.

To eliminate these sidelobes, we make an emission mask and CLEANed much deeper in the masked region, with $10^7$ CLEAN components per velocity channel, a gain parameter of 0.05, and only for positive emissions.
We use a signal-to-noise (S/N) ratio cube and make the mask by selecting the volumes (pixels) with at least one pixel with $>3.5\sigma$ and more than 32 pixels with $>2.5\sigma$, and extended their volumes to the pixels with $>1.5\sigma$.
The choice of these thresholds is arbitrary and is set by trial and error.

Reducing the negative sidelobes often brings up emissions that were previously hidden within the sidelobes.
They are not included in the first mask, and hence, we repeat the CLEAN process iteratively three times.
A mask is generated from the CLEANed cube in the previous iteration, starting from the CLEANed cube by \citet{Koda:2023aa}.
The final residual emission within the mask is $\sim 1/10\sigma$.
After the last iteration, we run CLEAN in the same way, but over the entire field without a mask nor the restriction for positive emissions.
This last step is to capture any missed emission at very low levels -- just in case it exists, but in practice it does not affect final emission maps.

The final data cube has grid sizes of 0.25" and $5\kmps$.
The native beam size and RMS noise are $2.10"\times 1.71"$ (PA=-88.0~deg) and 3.1~mJy/beam, respectively.
However, instead of using the native beam, we adopt a round 2.1" beam as a CLEAN beam (restoring beam) to match the resolution with that of the CO(2-1) data.

\subsection{CO(2-1)} \label{sec:CO21reduction}

The CO(2-1) data are from three ALMA projects (\#2013.1.01161.S, 2015.1.00121.S, and 2016.1.00386.S).
The visibility data are calibrated with the data reduction pipeline scripts from the observatory.
We exclude three execution blocks: two covering only the galactic center and one with short observing time and large noise.

\begin{figure}
    \centering
    \includegraphics[width=0.4\linewidth]{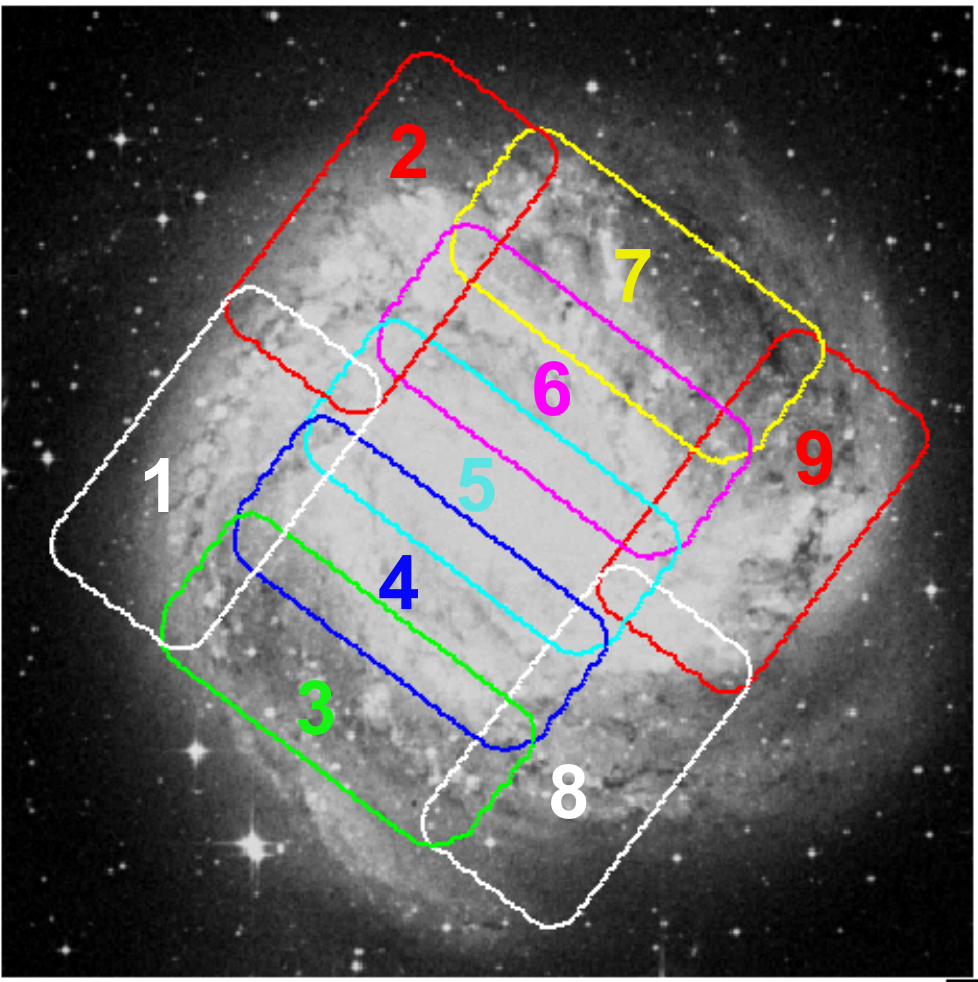}
    \caption{Nine fields observed separately in CO(2-1), overlaid on a $12'\times 12'$ optical image of M83 from the Digitized Sky Survey (DSS). Each of these fields has its own set of 12m, 7m, and TP observations.}
    \label{fig:co21fov}
\end{figure}

\subsubsection{Nine Subfields}

The whole field was divided into nine fields, each of which is observed separately (Figure \ref{fig:co21fov}).
The combined field is slightly smaller than that of the CO(1-0) observations that covered the whole molecular gas disk of M83  \citep[see ][]{Koda:2020aa, Koda:2023aa}.
Each field has a different $uv$ coverage and sensitivity.

To find a suitable beam size for their joint-imaging, we first image the nine fields separately.
We made a 12m+7m dirty data cube of each field with robust=0.0 in a limited velocity range for blank sky (only 5 channels of blank sky, starting from $250\kmps$ with a width of $5\kmps$)
and measured the beam size and RMS noise (Table \ref{tab:CO21data}).
The beam size at the full width at half maximum (FWHM) is often measured by fitting a Gaussian shape to dirty beam.
However, since the shapes of dirty beams can deviate from a Gaussian as it depends on the $uv$ distribution, the Gaussian fit does not fairly characterize the beam shapes for the comparison.
Therefore, we fit an ellipse to the FWHM contour of dirty beam.

We did not include the TP data for this comparison, because the FWHM contour of dirty beam is determined predominantly by the long baseline (12m+7m) data alone \citep{Koda:2019aa}.

\begin{deluxetable}{cccc}
\tablecaption{CO(2-1) 12m+7m Data\label{tab:CO21data}}
\tablewidth{0pt}
\tablehead{
\colhead{Field}   & \colhead{$b_{\rm maj}$, $b_{\rm min}$, $b_{\rm PA}$} & \colhead{RMS}  \\
\nocolhead{}        & \colhead{[", ", $\deg$]} & \colhead{[mJy/beam]} 
}
\startdata
1 & 0.94, 0.79, \,\,-89.2 &  4.2   \\
2 & 1.01, 0.81, +80.6 &  3.4   \\
3 & 1.75, 1.04, \,\,-91.3 &  3.6   \\
4 & 1.30, 0.94, +76.2 &  4.2   \\
5 & 1.10, 0.75, \,\,-92.4 &  3.7   \\
6 & 1.36, 0.98, +77.1 &  4.4   \\
7 & 2.06, 1.00, \,\,-77.2 &  3.4   \\
8 & 0.97, 0.80, +67.1 &  8.3   \\
9 & 0.79, 0.68, +81.0 &  6.1   \\
\enddata
\tablecomments{
We set robust=0.0 for 12m+7m imaging.
The RMS is from the first 5 channels starting from $250\kmps$ with a channel width of $5\kmps$.}
\end{deluxetable}

\subsubsection{TP Cube to Visibilities} 

The TP cube was converted to visibilities with TP2VIS.
The combination of the nine fields did not cover the entire molecular disk, and the significant emission outside the observed field is missed.
This total lack causes low-amplitude, but systematic, fringes near the edges of the combined field when the cube is deconvolved with the TP beam in TP2VIS.
To mitigate this problem, we filled the outside area with the CLEANed CO(1-0) data, by smoothing it to the CO(2-1) TP resolution (28.3") and by scaling it by 0.55 (a typical line ratio in brightness temperature near the edge).
Ten pixels (28.1") at the edge of the CO(2-1) data are used as a transition zone.
We used the ``$[1-\cos(\pi*(d/10))]/2$" function for the transition from CO(1-0) to CO(2-1) in brightness, where $d$ is the distance in pixel from the edge of the CO(2-1) TP data.
The transition zone is outside the field of view of the 12m+7m data (at a 50\% sensitivity).
Hence, the CO(2-1) emission within the final field of view practically comes from the CO(2-1) observations even after this operation.

\subsubsection{Imaging}

Table \ref{tab:CO21data} shows that the beam size varies by a factor of $\sim 2$ among the fields.
The largest beam size is 2.1" along the major axis (Field 7).
Hence, we target the resolution of 2.1" and will use the CLEAN (restore) beam size of 2.1".
In many fields, the beams are elongated significantly in the RA direction (the beam position angle $b_{\rm PA}$ around $\pm 90\deg$), and their major and minor axis ratios are about 2.
Thus, we set a Gaussian taper with a FWHM of 1.8" in the DEC direction in the $uv$ space along with robust=0.0 (as used for the measurements of Table \ref{tab:CO21data}) for a rounder beam shape.
The angular size of 2.1" corresponds to a baseline length of about 100~k$\lambda$,
and the $uv$ space within $<100$~k$\lambda$ is fairly filled for all the fields.
It also matches the resolution of the CO(1-0) data.

The imaging is done in two steps with MIRIAD.
First, we make a mask from the CLEANed CO(1-0) cube in the same way as discussed in Section \ref{sec:CO10reduction}.
We then use it to CLEAN the CO(2-1) data deeply with $2\times 10^7$ CLEAN components in each velocity channel.
The number of CLEAN components for CO(2-1) is set to be twice that for CO(1-0), because the areal ratio of their native dirty beams is about 2.
Second, we make another mask using the CLEANed CO(2-1) cube in the same way and add it to the CO(1-0) mask.
CLEAN is run again with the combined mask.
Although we run a final CLEAN without any mask for the CO(1-0) data,
we do not run CLEAN without mask for CO(2-1) because the sensitivity varies significantly among the fields (Table \ref{tab:CO21data}).

The grid sizes are 0.25" and $5\kmps$.
The native dirty beam size and RMS noise are 1.54"$\times$1.50" (PA=84.7~deg) and 4.1~mJy/beam (at the best parts of the data), respectively.
We adopt the CLEAN beam size of 2.1".

\section{Sensitivity Cuts for $R_{21/10}$ Analysis} \label{sec:senscut}

\begin{figure}
    \centering
    \includegraphics[width=0.6\linewidth]{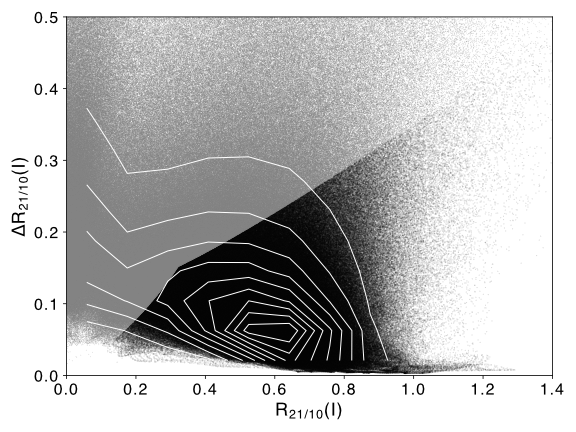}
    \caption{The sensitivity-based cuts in the $\Delta R_{21/10}$ vs. $R_{21/10}$ plane.
    The contours show the data density in this plot and are drawn at the 90, 80, 70, ..., 10\% levels of the peak density.}
    \label{fig:sencut}
\end{figure}

We use only the data of high significance for the $R_{21/10}$ analysis with the signal-to-noise (S/N) criteria of
$I_{10}/\Delta I_{10}>3$ and $I_{21}/\Delta I_{21}>3$.
The $I_{10}$ and $I_{21}$ are the integrated intensities in CO(1-0) and CO(2-1), and $\Delta I_{10}$ and $\Delta I_{21}$ are their associated errors, respectively.
In the first order, these errors propagate to an error in $R_{21/10}$ as
\begin{equation}
    \Delta R_{21/10}=R_{21/10}\sqrt{ \left( \frac{\Delta I_{10}}{I_{10}} \right)^2+ \left( \frac{\Delta I_{21}}{I_{21}} \right)^2}. \label{eq:R21error}
\end{equation}
Figure \ref{fig:sencut} shows the selection in the $\Delta R_{21/10}$-$R_{21/10}$ plane.
The black data points are used in our analysis, and the gray are removed.
The contours show the data density and are drawn at the 90, 80, ..., 10\% levels of the peak density.

Cursorily, the boundary between the black and gray can be understood in the following way.
The boundary line shows a kink around $R_{21/10}\sim$0.32.
The lower $R_{21/10}$ regions ($R_{21/10}\ll $0.32) typically have lower $I_{10}$ and $I_{21}$ (Figure \ref{fig:correlations}), and both $I_{10}$ and $I_{21}$ can approach our sensitivity limits ($I_{10}/\Delta I_{10}\sim 3$ and $I_{21}/\Delta I_{21}\sim3$).
Therefore, the boundary runs around $\Delta R_{21/10} \sim (\sqrt{2}/3) R_{21/10}$.
At higher $R_{21/10}$ ($\gg 0.32$), because of the high line ratio, the S/N ratio of $I_{21}$ tends to be higher than that of $I_{10}$ (i.e., $I_{21}/\Delta I_{21} \gg I_{10}/\Delta I_{10}$).
Therefore, the boundary approaches $\Delta R_{21/10}\approx R_{21/10} (\Delta I_{10}/I_{10})\sim (1/3) R_{21/10}$.
Obviously, $R_{21/10}\sim 0.32$ is the transition between those two regimes.

$\Delta R_{21/10}\sim$0.06 around the peak density (see the contours), which is sufficiently small for the discussions in this paper.
Even in the worst case (the boundary between the black and gray), $\Delta R_{21/10}\sim$0.25 around $R_{21/10}\sim$0.7.
This is still ok to separate the low and high ratio gas ($R_{21/10}<$0.7 and $>0.7$) in a statistical sense over the large areas of the molecular structures discussed in this paper.

\section{The HII-Region Mask} \label{sec:HIImask}
We generate a mask to separate the regions \textit{around} and \textit{outside} H$\alpha$-emitting regions, dubbed the HII-region mask, by detecting such regions in the HST images \citep{Blair:2014aa}.
We only aim to identify the areas under the influence of H$\alpha$ emission to a faint level.
Hence, while most objects we detect are individual HII regions, 
we do not attempt to generate their complete catalog.
We aim for a coarse photometric accuracy with a factor of $\sim$2 uncertainty.
Thus, we omit some corrections, e.g., for the contaminations of [NII] emissions, stellar absorptions, and other types of objects (e.g., AGB stars).
The photometric zero points are from \citet{Calamida:2022aa}.

We run SExtractor \citep{Bertin:1996lr} with the F657N (H$\alpha$) and F673N images: their difference, F657N-F673N, for detections, and F657N (or F673N) alone for measurements.
We use the SExtractor's default parameters except for the detection threshold of $3\sigma$ (instead of $1.5\sigma$ in the default) and minimum detection area of 10 pixels (about 3~pc diameter for a circle aperture).

The F673N band is adjacent to F657N in wavelength and nicely removes the background continuum emission.
However, F673N includes the [SII] emissions, which reduces H$\alpha$ fluxes in the detection image by about 20\%, a typical [SII]/H$\alpha$ flux ratio of HII regions \citep{Long:2022aa}.
This affects the detections only at the faintest level, which in practice is negligible given the faint flux limit achieved in the end (see below).

Figure \ref{fig:HIIselection}a compares the F657N and F673N fluxes, $F_{\rm F657N}$ and $F_{\rm F673N}$, of the objects detected in the area of Figure \ref{fig:HIImask}.
There is a distinct horizontal sequence along $\log F_{\rm F657N}/F_{\rm F673N}\sim0$ (i.e., no H$\alpha$ excess), which are clearly stars or star clusters in the images.
We remove them with a cutoff of $\log F_{\rm F657N}/F_{\rm F673N}<0.05$.
We do not attempt to remove other contaminants because they are likely rare.

SExtractor detects faint objects efficiently, but occasionally breaks up extended objects with internal structures (e.g., large HII regions) into smaller pieces \citep{Bautista:2023aa}.
We do not connect them together, since their whole areas are covered as an ensemble of the small pieces in the mask anyway.

Figure \ref{fig:HIIselection}b shows the H$\alpha$ luminosity ($L_{\rm H\alpha}$) function of the detected objects (with no extinction correction).
The shape is similar to that from a more sophisticated analysis in M51 \citep[][see their Figure 11]{Scoville:2001aa},
while we reach a fainter $L_{\rm H\alpha}$ due to twice a closer distance of M83.
Our luminosity function peaks around $L_{\rm H\alpha}\sim 10^{36}\rm \, erg/s$, and the detections reach down to $L_{\rm H\alpha}\sim 10^{35}\rm \, erg/s$.

As a guideline to assess the depth, the Orion Nebula (M42/M43) in the Milky Way is among the least impressive Galactic OB associations \citep{Hillenbrand:1997aa} and has $L_{\rm H\alpha}\sim 7\times 10^{36}\rm \, erg/s$ \citep[extinction-free; ][]{Scoville:2001aa}.
The analysis here would detect the Orion Nebula in M83 even under an extinction of $A_{\rm V}\sim$5-6~mag (note the extinction in H$\alpha$ wavelength is $A_{\rm H\alpha}\sim 0.8A_{\rm V}$).
In addition, HII regions are rarely completely-obscured to become undetectable in H$\alpha$ even with shallower H$\alpha$ imaging \citep{Hassani:2023aa}.

We include the regions around the detected objects in the mask.
Their diameters are set to 50, 100, and 200~pc for $L_{\rm H\alpha}<10^{37}$, $10^{37-38}$, and $>10^{38}\,\rm erg/s$.
These diameters cover, with comfortable margins, the whole extents of the HII regions identified in M51 by \citet[][their Figure 15]{Scoville:2001aa}.
Thus, they cover the areas under the influence of H$\alpha$ emission.
The smallest diameter of 50~pc is large for faint HII regions and is a round-up of the CO-data resolution of 46~pc.
This HII-region mask likely includes the areas beyond the actual extents of the HII regions.
The mask is presented in Figure \ref{fig:HIImask} as green circles.

\begin{figure}
    \centering
    \includegraphics[width=1.0\linewidth]{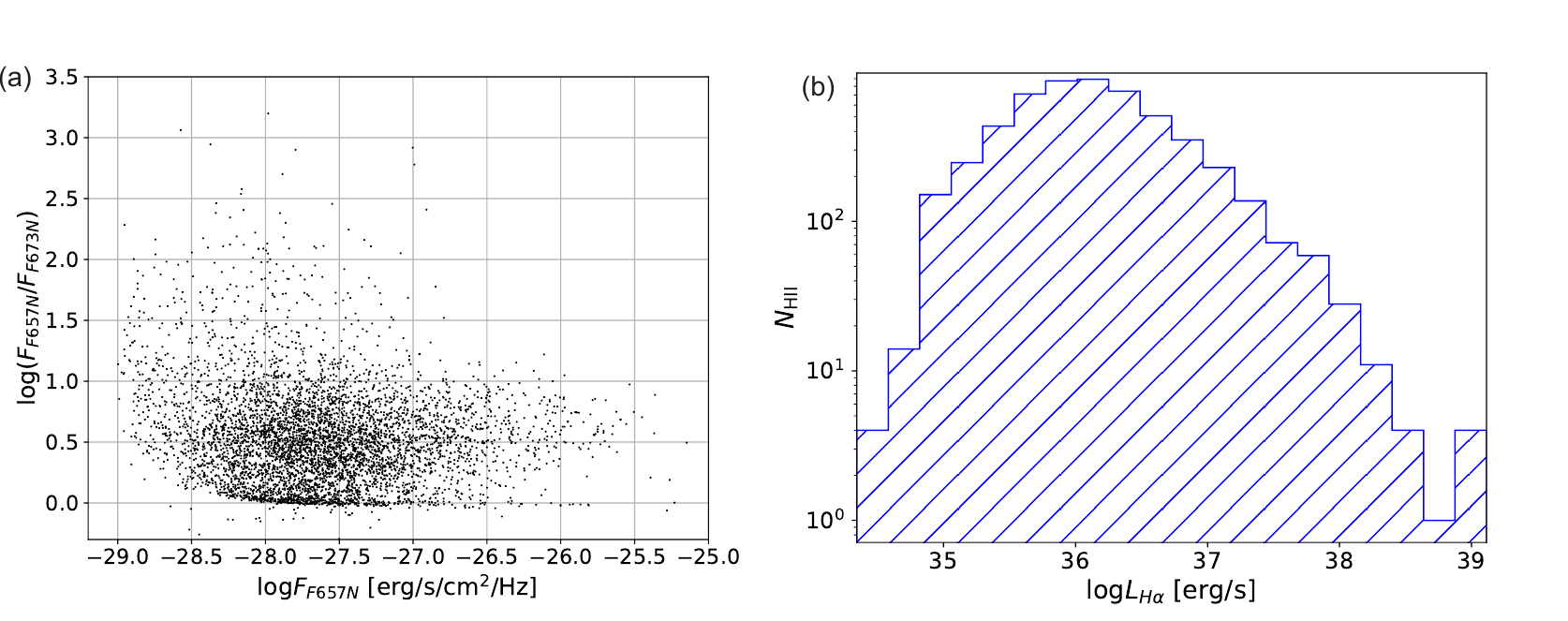}
    \caption{Plots for the HII region candidates detected in the archival HST images \citep{Blair:2014aa} in the area of Figures \ref{fig:dormant} and \ref{fig:HIImask}.
    (a) The F657N/F673N flux ratio vs F657N flux plot. The horizontal sequence around $\log F_{\rm F657N}/F_{\rm F673N}\sim0$ (no H$\alpha$ excess) consists of stars and star clusters and are removed from the mask.
    (b) The H$\alpha$ luminosity function of the detected objects (with no extinction correction).
    It peaks around $L_{\rm H\alpha}\sim 10^{36}\rm \, erg/s$.
    The detections reach as faint as $L_{\rm H\alpha}\sim 10^{35}\rm \, erg/s$.
    }
    \label{fig:HIIselection}
\end{figure}

\begin{acknowledgements}
JK thanks Dr. Keiichi Wada and his group members for hosting him at Kagoshima University while he is working on this paper.
We also thank the anonymous referee for valuable comments, especially for the analysis of Figures \ref{fig:HIImask} and \ref{fig:HIImask_hist}.
This work makes use of the following ALMA data: ADS/JAO.ALMA\#2017.1.00079.S, 2013.1.01161.S, 2015.1.00121.S, and 2016.1.00386.S.
ALMA is a partnership of ESO (representing its member states), NSF (USA) and NINS (Japan), together with NRC (Canada), MOST and ASIAA (Taiwan), and KASI (Republic of Korea), in cooperation with the Republic of Chile. The Joint ALMA Observatory is operated by ESO, AUI/NRAO and NAOJ.
The National Radio Astronomy Observatory is a facility of the National Science Foundation operated under cooperative agreement by Associated Universities, Inc..
JK acknowledges support from NSF through grants AST-2006600 and AST-2406608.
F.M. is supported by JSPS KAKENHI grant No. JP23K13142.

\end{acknowledgements}

\software{
CASA \citep{CASA-Team:2022aa},
MIRIAD \citep{Sault:1995kl, Sault:1996uq},
TP2VIS \citep{Koda:2019aa}}

\end{document}